\documentclass[prb,a4paper,twocolumn,superscriptaddress,longbibliography
]{revtex4-2}
\usepackage{multirow}
\usepackage{amssymb}
\usepackage{amsmath}
\usepackage{amsfonts}
\usepackage{graphicx}
\usepackage{bm}
\usepackage{xcolor,colortbl}
\usepackage{multirow}
\usepackage{float}
\usepackage{comment}
\usepackage[normalem]{ulem}
\usepackage{relsize}
\usepackage{cmap}
\usepackage{bbold}
\usepackage{physics}
\usepackage{setspace}
\usepackage{cancel}

\usepackage[colorlinks,citecolor=blue,linkcolor=red,urlcolor=blue]
{hyperref}

\definecolor{myGray}{gray}{0.9}

\usepackage[final]{pdfpages}
\usepackage{pgffor}

\makeatletter
\AtBeginDocument{\let\LS@rot\@undefined}
\makeatother

\DeclareMathOperator{\im}{Im}

\begin{document}

\title{Instability of the engineered dark state in two-band fermions under number-conserving dissipative dynamics}
		
\author{A. A. Lyublinskaya}	

\affiliation{\hbox{L.~D.~Landau Institute for Theoretical Physics, acad. Semenova av. 1-a, 142432 Chernogolovka, Russia}}

\affiliation{Department of Physics, HSE University, 101000 Moscow, Russia}

\author{P. A. Nosov}	

\affiliation{\hbox{Department of Physics, Harvard University, Cambridge, Massachusetts 02138, USA}}

\author{I. S. Burmistrov}	

\affiliation{\hbox{L.~D.~Landau Institute for Theoretical Physics, acad. Semenova av. 1-a, 142432 Chernogolovka, Russia}}

\affiliation{Laboratory for Condensed Matter Physics, HSE University, 101000 Moscow, Russia}

\date{\today} 
	
\begin{abstract}
Correlated quantum many-body states can be created and controlled by the dissipative protocols. 
Among these, particle number-conserving protocols are particularly appealing due to their ability to stabilize topologically nontrivial phases. Is there any fundamental limitation to their performance? We address this question by examining a general class of models involving a two-band fermion system subjected to dissipation designed to transfer fermions from the upper band to the lower band. 
By construction, these models have a guaranteed steady state -- a dark state -- with a completely filled lower band and an empty upper band. In the limit of weak dissipation, we derive equations governing the long-wavelength and long-time dynamics of the fermion densities and analyze them numerically. These equations belong to the Fisher-Kolmogorov-Petrovsky-Piskunov reaction-diffusion  universality class. Our analysis reveals that the engineered dark state is generically unstable, giving way to a new steady state with a finite density of particles in the upper band. We also estimate the minimum system sizes required to observe this instability in finite systems. Our results suggest  that number-conserving dissipative protocols may not be a reliable universal tool for stabilizing dark states.
\end{abstract}

\maketitle


\section{Introduction}

Open quantum many-body systems with dissipative dynamics is a topic of surging interest~\cite{Diehl2016,LeHur2018,Skinner2019,Rudner2020,Thompson2023,Diehl2023r}. The unique feature of this field lies in the possibility of using dissipative protocols for stabilizing correlated stationary states of matter that are inaccessible within equilibrium settings~\cite{Lechner2013,Piazza2014,Keeling2014,Altman2015,Devoret2015,Kollath2016}. Over the past few decades, numerous dissipation-induced states of matter and associated non-equilibrium phase transitions have been identified and intensely studied ~\cite{Diehl2008,Kraus2008,Verstraete2009,Weimer2010,Diehl2011,Tomadin2011,Bardyn2012,Bardyn2013,Otterbach2014,Koenig2014,Lang2015,Budich2015,Iemini2016,Zhou2017,Gong2017,Goldstein2019,Fisher2019a,Fisher2019b,Shavit2020,Tonielli2020,Yoshida2020,Gau2020a,Gau2020b,Bandyopadhyay2020,Santos2020,Altland2021,Beck2021,Nava2023,Shkolnik2023,Fava2023,Poboiko2023,Perfetto2023,Rowlands_2024,Gerbino2024,Poboiko2024,Yoshida2024,Kawabata2024}.

Dissipative protocols are typically described by the evolution of the density matrix $\rho$ governed by the Gorini-Kossakovski-Sudarshan-Lindblad (GKSL) master equation \cite{Lindblad1976,GKS1976}. The most widely-studied dissipative protocols are designed in a particle non-conserving way, so that the engineered steady state contains fewer particles than the initial state. Technically, this corresponds to jump operators in the GKSL equation that are linear in particle creation and annihilation operators, which significantly simplifies the analysis of such dynamics~\cite{Diehl2016,Thompson2023,Prosen2008,Prosen2010}.

The dissipative evolution with $U(1)$ particle conservation has lately gained a lot of attention~\cite{ Diehl2023r}, mainly due to its ability to stabilize topologically nontrivial steady states characterized by topological gauge field responses. However, the description of particle kinetics and relaxation under such dissipation is significantly complicated 
due to the jump operators that are bilinears in particle creation and annihilation operators, making the resulting models strongly-interacting. The same challenge emerges in systems with projective measurements~\cite{Fava2023,Poboiko2023,Poboiko2024}.

\begin{figure}[b!]
\centerline{\includegraphics[width=0.79\columnwidth]{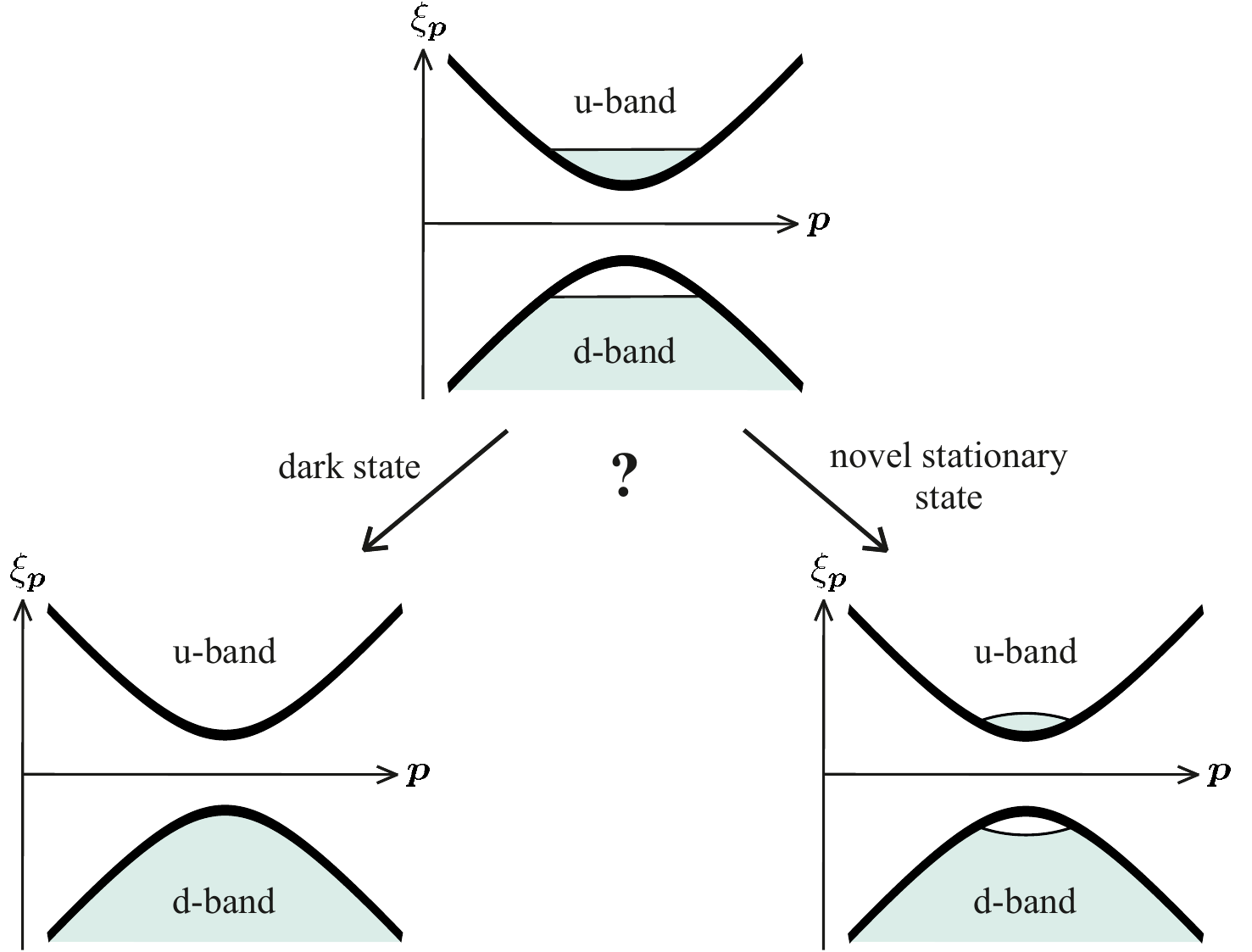}
}
\caption{Possible directions of initial state evolution in the two-band fermionic dissipative system. Contrary to naive expectations, the evolution follows the right path, resulting in a stationary state with a finite population in the upper band.}
\label{Fig:intro}
\end{figure}

Recently, significant progress has been made in understanding $U(1)$-conserving protocols. 
A model of two-band fermions with dissipation aimed at stabilizing a topological state by emptying one band and populating the other was proposed~\cite{Tonielli2020}. Further analysis~\cite{Nosov2023} revealed that the density in each band follows the Fisher-Kolmogorov-Petrovsky-Piskunov (FKPP) equation \cite{Fisher1937,Kolmogorov1937}, which describes various dynamical processes, from the propagation of combustion fronts to bacterial spreading \cite{FKPP1988,FKPP2,FKPP3,Aleiner2016,Zhou2023}. The FKPP equation indicates diffusive spreading of density at intermediate length/time scales and ballistic front propagation at long scales. Consequently, the dark state (DS) with a completely empty upper band is unstable, leading to a new steady state with nonzero density in the upper band (Fig. \ref{Fig:intro}).

This leads to a fundamental question: {\it Is the engineered dark state, characterized by a completely empty upper band, stable in a generic two-band fermionic model with particle-number-conserving dissipation?} In this paper, we address this question for a class of models introduced in
Ref.
~\cite{Lyublinskaya2023}, which generalizes the original model by Tonielli et al. ~\cite{Tonielli2020}. These two-band fermionic models are designed to have a 
DS with a completely empty upper band. However, we demonstrate that, generally, under weak dissipation rates, this engineered 
DS is unstable. The system evolves towards 
new steady state with a nonzero population in the upper band, and thus, the scenario depicted on the right side of Fig.~\ref{Fig:intro} is realized. We derive equations governing the evolution of the particle densities in each band, which generalize the FKPP equation. We also establish the physical meaning of all processes that contribute to these equations and identify their analogs in semiconductor theory. Our results suggest a potential failure of particle number-conserving dissipative protocols as a universal tool for stabilization of DS.

The outline of the paper is as follows. We start from the model formulation in terms of the GKSL master equation and Keldysh field theory in Sec. \ref{Sec:Model}. In Sec. \ref{Sec:SCBA&diffuson} we 
review the self-consistent Born approximation for the single-particle Green's function and the ladder approximation for the two-particle irreducible 
density-density correlation function 
(the so-called diffuson). After this, the main results of this work are described: different contributions to 
the diffuson's self-energy (Sec. \ref{Sec:SE}) and the generalized  FKPP equations which govern 
the dynamics of the particle density  
  (Sec.~\ref{Sec:FKPP}). We end the paper with discussions and summary in Sec.~\ref{Sec:Discussions} and Sec.~\ref{Sec:Summary}, respectively. Details of lengthy calculations are given in Appendices.

\section{Model\label{Sec:Model}}

\subsection{The GKSL equation}
 
\begin{figure}[t]
\centerline{\includegraphics[width=0.75\columnwidth]{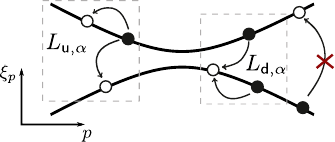}
}
\caption{Schematic representation of the fermion spectrum and the action of the jump operators $L_{\textsf{u},\alpha}$ and $L_{\textsf{d},\alpha}$. A single jump operator can either change the state of a particle within one band or transfer a particle from the upper band to the lower one.}
\label{Fig:model}
\end{figure}

We consider the dissipative dynamics of two-band fermions governed by the GKSL master equation \cite{Lyublinskaya2023},
\begin{gather}
\frac{d\rho}{dt} {=} \int\limits_{\bm{x}} 
\Bigl ( i [\rho,H_{\rm 0}] {+}\sum_{j{=}\{\textsf{a},\alpha\}} \gamma_j
\bigl (2 L_j \rho L_j^\dag 
{-} \{L_j^\dag L_j, \rho\}\bigr )\Bigr ) ,
\label{eq:GKSL}
\end{gather}
where $\textsf{a}{=}\textsf{u},\textsf{d}$ and $\alpha{=}\uparrow,\downarrow$. The jump operators are given as local-in-space bilinears in terms of fermionic operators: 
\begin{equation}
    L_{\textsf{u},\alpha} {=} \psi^{\dag}_{\alpha}(\bm{x})  l_{\textsf{u}}(\bm{x}),\quad L_{\textsf{d},\alpha} {=} \psi_{\alpha}(\bm{x}) l^{\dag}_{\textsf{d}}(\bm{x})\;.
\end{equation}
The unitary part of the dynamics is governed by the translationally invariant Hamiltonian density
\begin{equation}
    H_0 {=} \sum\limits_{\alpha\beta}\psi^{\dag}_{\alpha}(\bm{x}) \hat{h}_{\alpha\beta}\psi_{\beta}(\bm{x}),\quad \hat{h}(\bm{p}){=}\xi_p U_{\bm{p}} \sigma_z U^\dag_{\bm{p}}\;.
\end{equation}
Here $\xi_p$ is an auxiliary non-negative function of the momentum $p{=}|\bm{p}|$ and $\sigma_z{=}{\rm diag}\{1,{-}1\}$ is the standard Pauli matrix. $U_{\bm{p}}$ is an arbitrary momentum dependent $2{\times}2$ unitary matrix which can be used to rotate fermionic operators from the spin to band basis $c_{\bm{p}} {=} \{c_{\bm{p},\textsf{u}},c_{\bm{p},\textsf{d}}\}^{\texttt{T}}{=} U^\dag_{\bm{p}} \psi_{\bm{p}}$ and, thus, to make the single particle Hamiltonian diagonal, $\xi_p \sigma_z$. The jump operators 
involve another set of the fermionic operators: 
$l_{\bm{p},\textsf{a}}{=} v_{\bm{p}} c_{\bm{p},\textsf{a}}$, where  $v_{\bm{p}}$ is an auxiliary complex function of momentum with $|v_{\bm{p}}|{\neq} 0$. Parameters $\gamma_{\textsf{a},\alpha}{>}0$ determine the rate of dissipation in the model~\footnote{Our model is a generalization of the model introduced originally in~\cite{Tonielli2020} for which  $U_{\bm{p}}{=}(p_x {-} ip_y\sigma_z{-}i m \sigma_y)/\sqrt{d_p}$, $v_p{=}\sqrt{d_p}$,  and $d_p{=}\xi_p{=}p^2{+}m^2$.}. 
At half-filling, Eq.~\eqref{eq:GKSL} has a steady state solution, the 
DS, $\rho{=}|D\rangle \langle D|$, in which the $\textsf{d}$-band is fully occupied, while the $\textsf{u}$-band is empty: $c^\dag_{\bm{p},\textsf{d}}|D\rangle{=}c_{\bm{p},\textsf{u}}|D\rangle{=}0$. This is ensured as $L_{\textsf{u}/\textsf{d},\alpha}$ cannot transfer particles from the $\textsf{d}$-band to the $\textsf{u}$-band (see Fig.~\ref{Fig:model}).

\subsection{Keldysh action}

The GSKL equation can be written as a path integral on the Keldysh contour \cite{Diehl2016,Thompson2023} with the following generating function ($\int_{\bm{p},t}{\equiv}\int dt\int d^d\bm{p}/(2\pi)^d$)
\begin{align}
Z & {=} \! \int \! \mathcal{D}[\overline{c},c]\,  e^{i S_{\rm 0} {+} i S_{\rm L}},  \, S_{\rm 0}
{=}\int\limits_{\bm{p},t}
\sum_{s=\pm} s\,
\overline{c}_{\bm{p},s} \bigl(  i \partial_t {-} \xi_{\bm{p}}\sigma_z \bigr )  c_{\bm{p},s} , \notag \\
S_{\rm L} & {=}   i (2\pi)^d \!\!\int\limits_{\bm{p_{j}},t}\! \delta(\bm{p_1}{-}\bm{p_2}{+}\bm{p_3}{-}\bm{p_4}) 
\sum_{s,s^\prime{=}\pm} \hspace{-0.1em}(
\delta_{ss^\prime}{-}2\delta_{s,{-}}\delta_{s^\prime,{+}})
\notag \\
 &\times \sum_{\textsf{a},\alpha}  \gamma_{\textsf{a},\alpha}(\overline{c}_{\bm{p_1},s}[\mathcal{L}^{(\textsf{a},\alpha)}_{\bm{p_2p_1}}]^\dag 
c_{\bm{p_2},s})
(\overline{c}_{\bm{p_3},s^\prime}\mathcal{L}^{(\textsf{a},\alpha)}_{\bm{p_3p_4}}
c_{\bm{p_4},s^\prime})
.
\label{eq:SL:ud}
\end{align}
Here $d$ is the spatial dimensionality, $c_{\bm{p},\textsf{a}}{=}\{c_{\bm{p},\textsf{a},+},c_{\bm{p},\textsf{a},-}\}^{\texttt{T}}$ and $\bar{c}_{\bm{p},\textsf{a}}{=}\{\bar{c}_{\bm{p},\textsf{a},+},\bar{c}_{\bm{p},\textsf{a},-}\}$ are 
Grassmanian fields on the forward/backward parts of the Keldysh contour. They correspond to the annihilation $c_{\bm{p},\textsf{a}}$ and creation $c^\dag_{\bm{p},\textsf{a}}$ operators, respectively. 
The action $S_L$, which represents the dissipative part of the GSKL equation, involves four fermionic operators at the same time $t$, and thus, formally resembles instantaneous interactions relaxing fermions'
momentum and energy. However, we note that this dissipative interaction couples different branches of the Keldysh contour and requires a particular equal-time regularization \cite{Diehl2016,Tonielli2020,Nosov2023,Lyublinskaya2023}.
The four matrices $\mathcal{L}^{(\textsf{a},\alpha)}$ act in the band space and are given as
\begin{equation}
[\mathcal{L}^{(\textsf{u},\alpha)}_{\bm{pq}}]_{\textsf{ab}}{=}v_{\bm{q}} [U_{\bm{p}}^\dag]_{\textsf{a},\alpha}\delta_{\textsf{bu}}, \,\,  [\mathcal{L}^{(\textsf{d},\alpha)}_{\bm{pq}}]_{\textsf{ab}} {=} {-}v_{\bm{p}}^* [U_{\bm{q}}]_{\alpha,\textsf{b}}\delta_{\textsf{ad}}  .
\label{eq:A:matrices:0}
\end{equation}

We note that $S_{\rm L}$ can be thought of as arising from the scattering of fermions from bosons, after taking the trace over the bosonic degrees of freedom \cite{Lyublinskaya2023}.
The conservation of the total number of fermions in the model is reflected in the strong $U(1)$ symmetry of the action $S_{\rm 0}{+}S_{\rm L}$. Translation invariance, however, is only a weak symmetry of $S_{\rm 0}{+}S_{\rm L}$. For detailed discussion of weak and strong symmetries 
see Refs.~\cite{Diehl2016,Buca2012,Albert2014,Diehl2023r}.

\begin{figure}[t]
\centerline{\includegraphics[width=1\columnwidth]{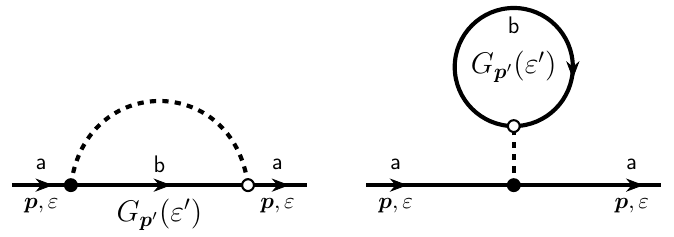}
}
\caption{Self-energy diagrams of the Fock- and Hartree-type in the self-consistent Born approximation. The solid lines denote the self-consistent Green's function. The dashed curve indicates the dissipation-induced interaction.}
\label{Fig:SCBA}
\end{figure}

\section{Self-consistent Born approximation and diffusion\label{Sec:SCBA&diffuson}}

\subsection{Self-consistent Born approximation}

In terms of the Keldysh action $S_{\rm 0}{+}S_{\rm L}$ existence of the 
DS 
can be seen from the structure of the single-particle Green's functions ${-}i\langle c_{\bm{q}}(t_1) \bar{c}_{\bm{q}}(t_2)\rangle$ computed in  a self-consistent manner from the Dyson equation in the lowest order in the dissipation rate $\gamma_{\textsf{a},\alpha}$ (see Fig.~\ref{Fig:SCBA}). These Green's functions  are diagonal in the $\textsf{u/d}$-space and after Keldysh rotation \cite{Kamenev2009} are given as \cite{Tonielli2020,Nosov2023,Lyublinskaya2023}
\begin{equation}
[\mathcal{G}^{R/A}_{\bm{q},\textsf{a}}(\varepsilon)]^{{-}1}{=}
\varepsilon{-}\xi_q s_\textsf{a} {\pm} i \bar{\gamma}_\textsf{a} |v_q|^2, \,\,
\mathcal{G}^K_{\bm{q},\textsf{a}}(\varepsilon) {=} 2 i s_\textsf{a} \im \mathcal{G}^{R}_{\bm{q},\textsf{a}}(\varepsilon) ,
\label{eq:G:SCBA}
\end{equation}
where $\bar{\gamma}_{\textsf{a}}{=} \int_p [U^\dag_{\bm{p}} \hat\gamma_{\textsf{a}} U_{\bm{p}}]_{\textsf{a}\textsf{a}}$ and $\hat\gamma_{\textsf{a}} {=} {\rm diag}\{\gamma_{\textsf{a},\uparrow},\gamma_{\textsf{a},\downarrow}\}$. The factor $s_\textsf{a}{=}{\pm} 1$ determines the distribution function $(1{-}s_\textsf{a})/2$ of the $c$-fermions in the $\textsf{a}$-band. Thus, Eq.~\eqref{eq:G:SCBA} describes the DS
with a completely empty $\textsf{u}$-band.

\begin{figure}[b]
\centerline{\includegraphics[width=1.0\columnwidth]{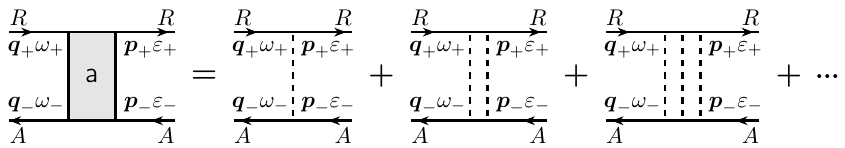}} 
\caption{The sum of ladder diagrams, $\textsf{a}$-diffuson.}
\label{Figure:Diffuson}
\end{figure}

\subsection{Diffuson's ladder}

However, Green's functions \eqref{eq:G:SCBA} do not characterize the spreading of the particle density through the system. The latter is characterized by the density-density correlation function, i.e. 
by the two-particle irreducible average  
\begin{equation}
\langle\!\langle c_{\bm{q}_-,\textsf{b},\beta}(\omega_-)  \bar{c}_{\bm{q}_+,\textsf{a},\alpha}(\omega_+){\cdot}c_{\bm{p}_+,\textsf{a}',\alpha'}(\varepsilon_+) \bar{c}_{\bm{p}_-,\textsf{b}',\beta'}(\varepsilon_-)\rangle\!\rangle
\end{equation}
with $\bm{p}_\pm {=} \bm{p}{\pm} \bm{Q}/2$, $\bm{q}{=}\bm{q}{\pm}\bm{Q}/2$, $\omega_\pm{=}\omega{\pm}\Omega/2$, $\varepsilon_\pm{=}\varepsilon{\pm} \Omega/2$. In the previous work \cite{Lyublinskaya2023} we calculated it in the ladder approximation, that is, we calculated the diffuson's diagram (see Fig.~\ref{Figure:Diffuson}). Obtained expression is $D_{\textsf{a}}^{(0)}(\bm{Q},\Omega, \bm{q},\omega,\bm{p},\varepsilon)\big[\Lambda_{\textsf{a}}^{(+)}\big]^{\alpha \alpha^{\prime}}\big[\Lambda_{\textsf{a}}^{(-)}\big]^{\beta^{\prime} \beta}\delta_{\textsf{a}=\textsf{a}'=\textsf{b}=\textsf{b}'}$, where
\begin{gather}
D_{\textsf{a}}^{(0)}(\bm{Q},\Omega, \bm{q},\omega,\bm{p},\varepsilon)=\frac{2v_{\bm{p}_{+}}v_{\bm{p}_{-}}^*}{1{-}f_{\textsf{a}}(\bm{Q}, \Omega)}\left[U_{\bm{q}_{+}}^{\dagger} \hat{\gamma}^{(\textsf{a})}U_{\bm{q}_{-}}\right]_{\textsf{aa}}\notag\\
{\times}\mathcal{G}_{\bm{q}_{+}, \textsf{a}}^R(\omega_{+}) \mathcal{G}_{\bm{q}_{-}, \textsf{a}}^A(\omega_{-})\mathcal{G}_{\bm{p}_{+}, \textsf{a}}^R(\varepsilon_{+}) \mathcal{G}_{\bm{p}_{-},\textsf{a}}^A(\varepsilon_{-}) .
\label{SI:Diffuson}
\end{gather}
 Here we introduced the function 
\begin{equation}
f_{\textsf{a}}(\bm{Q}, \Omega){=}\int_{\bm{k}} \frac{2 i v_{\bm{k}_{+}}v_{\bm{k}_{-}}^*\big[U_{\bm{k}_{+}}^{\dagger} \hat{\gamma}^{(\textsf{a})} U_{\bm{k}_{-}}\big]_{\textsf{aa}}}{\Omega{-}s_{\textsf{a}} (\xi_{\bm{k}_{+}}{-} \xi_{\bm{k}_{-}}){+}i\bar{\gamma}_{\textsf{a}}\big(|v_{\bm{k}_{+}}|^2{+}|v_{\bm{k}_{-}}|^2\big)}
\end{equation}
and two projectors on retarded and advanced spaces
\begin{equation}
\Lambda^{({+})}_\textsf{a}=\begin{pmatrix}
1 & s_\textsf{a} \\
0 & 0 
\end{pmatrix} , \quad 
\Lambda^{(-)}_\textsf{a}=\begin{pmatrix}
0 & -s_\textsf{a} \\
0 & 1 
\end{pmatrix}.\label{eq:Lambda}
\end{equation}
Here $s_\textsf{u}{=}{-}s_\textsf{d}{=}1$. Such two-particle correlation function computed in the ladder approximation has a canonical-type diffusion pole for $Q, \Omega {\to} 0$:
\begin{equation}
\frac{1}{1{-}f_{\textsf{a}}(\bm{Q},\Omega)}\simeq\frac{2\bar{\gamma}_{\textsf{a}}^2/\int_{\bm{k}}[U_{\bm{k}}^{\dagger}\hat{\gamma}^{(\textsf{a})}U_{\bm{k}}]_{\textsf{aa}}\left|v_{\bm{k}}\right|^{-2}}{D_{jl}^{(\textsf{a})} Q_j Q_l{-}i \Omega}.
\end{equation}

The matrix of diffusion coefficients has different contributions due to curvature of the spectrum $\xi_k$ and the function $v_k$. Additionally, the non-Abelian vector potential in the momentum space (Berry connection), $\mathcal{A}_j{=}i U^\dag_{\bm{k}} \partial_{\bm{k_j}} U_{\bm{k}}$, contributes to $D^{(\textsf{a})}_{jl}$ \cite{Lyublinskaya2023}. In the limit $\bar{\gamma}_{\textsf{a}}{\ll}1$ the matrix of diffusion coefficients reads 
\begin{equation}
D^{(\textsf{a})}_{jl} =\delta_{jl}\frac{\int_{\bm{p}} [U^\dag_{\bm{p}} \hat\gamma_{\textsf{a}} U_{\bm{p}}]_{\textsf{a}\textsf{a}}|v_p|^{-4} (\nabla_{\bm{p}} \xi_p)^2}{2\bar{\gamma}_{\textsf{a}} d \int_{\bm{k}} [U^\dag_{\bm{k}} \hat\gamma_{\textsf{a}} U_{\bm{k}}]_{\textsf{a}\textsf{a}}|v_k|^{-2}}
+O(1) .
\label{eq:res:D}
\end{equation}
In the ladder approximation there is intra-band diffusion only; the inter-band excitations do not diffuse \cite{Nosov2023,Lyublinskaya2023}. Due to their resemblance to analogous excitations in disordered electronic systems \cite{Kamenev2009}, we refer to these diffusive particle-hole modes as ``diffusons''.

\section{Diffuson's self-energy\label{Sec:SE}}

\begin{figure}[b]
\centerline{\includegraphics[width=1\columnwidth]{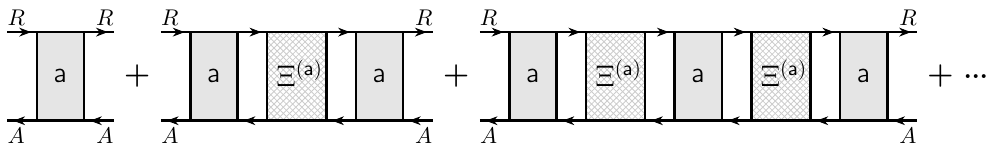}} 
\caption{General scheme of inserting the total self-energy $\Xi^{(\textsf{a})}$ into a $\textsf{a}$-diffuson's ladder with subsequent resummation of the ladder series.}
\label{Figure:SE:resummation}
\end{figure}

\subsection{Insertion of self-energy into the diffuson and resummation of the ladder series}

Although the GSKL dynamics \eqref{eq:GKSL} conserves the total number of particles, the number of fermions in each band separately is not conserved. Therefore, the diffusion poles in two-point correlation functions are not protected beyond the ladder approximation. Therefore, we consider the diffuson's self-energy. Let's denote the sum of inserted diagrams as $\xi^{(\textsf{a})}(\bm{Q},\Omega,\bm{k},\epsilon, \bm{k}',\epsilon')$. One insertion between two diffusons will result in an additional multiplier:
\begin{widetext}
\begin{equation}
\begin{gathered}
\int_{\epsilon,\bm{k},\epsilon',\bm{k}'}D_{\textsf{a}}^{(0)}(\bm{Q},\Omega,\bm{q},\omega, \bm{k},\epsilon)\big[\Lambda_{\textsf{a}}^{(+)}\big]^{\alpha\delta'}\big[\Lambda_{\textsf{a}}^{(-)}\big]^{\gamma'\beta}\xi^{(\textsf{a})}(\bm{Q},\Omega,\bm{k},\epsilon,\bm{k}',\epsilon')\big[\Lambda_{\textsf{a}}^{(+)}\big]^{\delta'\delta}\big[\Lambda_{\textsf{a}}^{(-)}\big]^{\gamma\gamma'}\\
\times D_{\textsf{a}}^{(0)}(\bm{Q},\Omega,\bm{k}',\epsilon',\bm{p},\varepsilon)\big[\Lambda_{\textsf{a}}^{(+)}\big]^{\delta\alpha'}\big[\Lambda_{\textsf{a}}^{(-)}\big]^{\beta'\gamma}=D_{\textsf{a}}^{(0)}(\bm{Q},\Omega, \bm{q},\omega,\bm{p},\varepsilon)\left[\Lambda_{\textsf{a}}^{(+)}\right]^{\alpha\alpha'}\left[\Lambda_{\textsf{a}}^{(-)}\right]^{\beta'\beta}\frac{2}{1-f_{\textsf{a}}(\bm{Q}, \Omega)}\\
\times\int_{\epsilon,\bm{k},\epsilon',\bm{k}'}v_{\bm{k}_{+}}v_{\bm{k}_{-}}^*\left[U_{\bm{k}'_{+}}^{\dagger} \hat{\gamma}^{(\textsf{a})}U_{\bm{k}'_{-}}\right]_{\textsf{aa}}\mathcal{G}_{\bm{k}_{+}, \textsf{a}}^R(\epsilon_{+}) \mathcal{G}_{\bm{k}_{-}, \textsf{a}}^A(\epsilon_{-}) \mathcal{G}_{\bm{k'}_{+}, \textsf{a}}^R(\epsilon'_{+}) \mathcal{G}_{\bm{k'}_{-},\textsf{a}}^A(\epsilon'_{-})\xi^{(\textsf{a})}(\bm{Q},\Omega,\bm{k},\epsilon,\bm{k}',\epsilon').
\end{gathered}
\end{equation}
\end{widetext}
For brevity, let's introduce the notation
\begin{equation}
\begin{gathered}
\Sigma^{(\textsf{a})}(\bm{Q},\Omega)=\int_{\epsilon,\bm{k},\epsilon',\bm{k}'}v_{\bm{k}_{+}}v_{\bm{k}_{-}}^*\big[U_{\bm{k}'_{+}}^{\dagger} \hat{\gamma}^{(\textsf{a})}U_{\bm{k}'_{-}}\big]_{\textsf{aa}}\mathcal{G}_{\bm{k}_{+}, \textsf{a}}^R(\epsilon_{+})\\
{\times} \mathcal{G}_{\bm{k}_{-}, \textsf{a}}^A(\epsilon_{-}) \mathcal{G}_{\bm{k'}_{+}, \textsf{a}}^R(\epsilon'_{+}) \mathcal{G}_{\bm{k'}_{-},\textsf{a}}^A(\epsilon'_{-})\xi^{(\textsf{a})}(\bm{Q},\Omega,\bm{k},\epsilon,\bm{k}',\epsilon').
\end{gathered}
\label{SI:Sigma}
\end{equation}

Therefore, the ladder series with self-energy insertions $D_{\textsf{a}}{=}D_{\textsf{a}}^{(0)}{+}D_{\textsf{a}}^{(0)}\xi^{(\textsf{a})}D_{\textsf{a}}^{(0)}{+}D_{\textsf{a}}^{(0)}\xi^{(\textsf{a})}D_{\textsf{a}}^{(0)}\xi^{(\textsf{a})}D_{\textsf{a}}^{(0)}{+}...$ results in resummation (see Fig.~\ref{Figure:SE:resummation})
\begin{equation}
\begin{gathered}
\frac{D_{\textsf{a}}(\bm{Q},\Omega,\bm{p},\varepsilon, \bm{q},\omega)}{D_{\textsf{a}}^{(0)}(\bm{Q},\Omega,\bm{p},\varepsilon, \bm{q},\omega)}=\sum\limits_{n=0}^{+\infty}\left[\frac{2\Sigma^{(\textsf{a})}(\bm{Q},\Omega)}{1{-}f_{\textsf{a}}(\bm{Q}, \Omega)}\right]^n=\\
=\frac{1{-}f_{\textsf{a}}(\bm{Q}, \Omega)}{1{-}f_{\textsf{a}}(\bm{Q}, \Omega){-}2\Sigma^{(\textsf{a})}(\bm{Q},\Omega)}.
\end{gathered}
\end{equation}

Here we see that self-energy turns into an additional term in the diffusive denominator, therefore
\begin{gather}
D_{jl}^{(\textsf{a})} Q_j Q_l{-}i \Omega\rightarrow D_{jl}^{(\textsf{a})} Q_j Q_l{-}i \Omega{+}\Xi^{(\textsf{a})}(0,0),\notag\\
\Xi^{(\textsf{a})}(0,0)={-}\frac{4\bar{\gamma}_{\textsf{a}}^2\Sigma^{(\textsf{a})}(0,0)}{\int_{\bm{k}}[U_{\bm{k}}^{\dagger}\hat{\gamma}^{(\textsf{a})}U_{\bm{k}}]_{\textsf{aa}}\left|v_{\bm{k}}\right|^{-2}}.
\label{SI:Xi}
\end{gather}
In what follows, we will compute the contributions to the diffuson's self-energy $\Xi^{(\textsf{a})}(0,0)$ which are of lowest order either in $\bar{\gamma}_{\textsf{a}}$ or in the deviation of fermion density from the one in the DS.

\begin{figure}[b]
\centerline{\includegraphics[width=1\columnwidth]{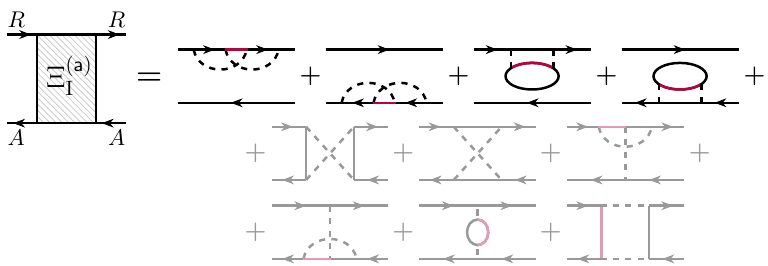}} 
\caption{The first contribution to the self-energy $\Xi^{(\textsf{a})}_{\rm I}$ corresponds to the process of impact ionization and is obtained from the insertion into the diffusion ladder of sections of the second order in the dissipation rate. Red lines denote the Green's function from the other ($\bar{\textsf{a}}$) band.  Diagrams that could have made a contribution, but are canceled out with each other or turned out to be equal to zero (at $\Omega,Q{\to}0)$, are shown in pale color.}
\label{Figure:SE:diagrams1}
\end{figure}

\begin{figure}[b]
\centerline{\includegraphics[width=0.9\columnwidth]{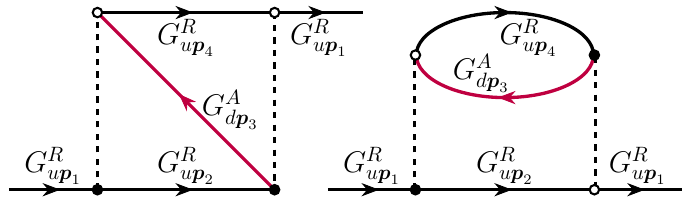}}

\,

\centerline{\includegraphics[width=0.5\columnwidth]{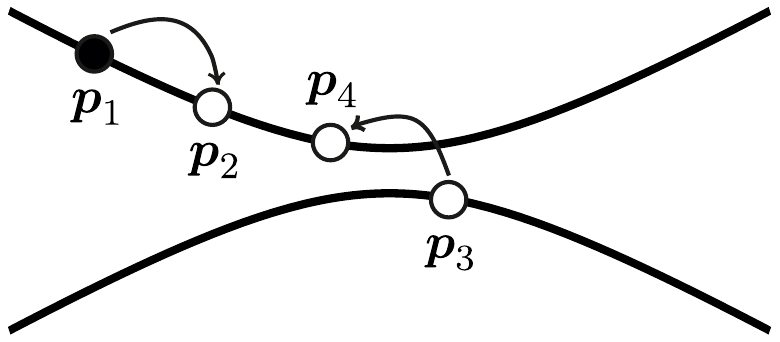}
}
\caption{Diagrams describing the process of impact ionization, presented in the form of a second-order contribution to the self-energy of the single-particle Green's function, $\Sigma^{R,(2)}_{\bm{p},\textsf{a}}(\varepsilon)$, and the corresponding many-particle process.}
\label{Figure:SE:FGR}
\end{figure}

\subsection{Impact ionization\label{SubSec:ImpIon}}

Focusing on the limit $\bar{\gamma}_\textsf{a}{\to}0$, we compute below all contributions to the diffuson self-energy $\Xi^{(\textsf{a})}(0,0)$ in the second order in $\bar{\gamma}_\textsf{a}$. We start from contributions depicted in Fig.~\ref{Figure:SE:diagrams1} that can be inserted between diffusive ladders in the same band. There are ten possible diagrams,  the sum of which we denote $\Xi^{(\textsf{a})}_{\rm I}$. Fortunately, there are drastic simplifications at $Q{=}\Omega{=}0$ for  $\Xi^{(\textsf{a})}_{\rm I}(0,0)$: 
diagrams shown in pale color cancel each other or vanish. The remaining four diagrams for $\Xi^{(\textsf{a})}_{\rm I}(0,0){\equiv}1/\tau_{\textsf{a}}$ yield (see Appendices~\ref{App:SE}~and~\ref{App:Detailed})
\begin{gather}
\frac{1}{\tau_{\textsf{a}}}{=}{-}\pi (2\pi)^d \int_{\bm{p}_i}
\delta(\bm{p}_1{-}\bm{p}_2{+}\bm{p}_3{-}\bm{p}_4) \Delta_{\gamma}(\bm{p}_1,\bm{p}_2,\bm{p_3},\bm{p_4})
\notag \\
\times
\frac{\left[U^{\dag}_{\bm{p}_1}\hat{\gamma}_{\textsf{a}} U_{\bm{p}_1}\right]_{\textsf{a}\textsf{a}}
\left|\left(v_{\bm{p}_4} U^{\dag}_{\bm{p}_3}\hat{\gamma}_{\textsf{a}} U_{\bm{p}_2}{-}v_{\bm{p}_2} U^{\dag}_{\bm{p}_3}\hat{\gamma}_{\textsf{a}} U_{\bm{p}_4}\right)_{\bar{\textsf{a}}\textsf{a}}\right|^2}
{\int_{\bm{k}}[U_{\bm{k}}^{\dagger} \hat{\gamma}_{\textsf{a}} U_{\bm{k}}]_{\textsf{aa}}\left|v_{\bm{k}}\right|^{{-}2}}
 ,
 \label{eq:res:tau}
 \end{gather}
 where a broadened Dirac delta function is introduced
 \begin{align}
 \Delta_{\gamma}(\bm{p}_1,\bm{p}_2,\bm{p_3},\bm{p_4}) & = \frac{1}{\pi}\Im \Bigl [\xi_{\bm{p}_1}{-}\xi_{\bm{p}_2}{-}\xi_{\bm{p}_3}{-}\xi_{\bm{p}_4} \notag \\
 -i(\overline{\gamma}_{\textsf{a}}|v_{\bm{p}_1}|^2&{+}\overline{\gamma}_{\textsf{a}}|v_{\bm{p}_2}|^2{+}\overline{\gamma}_{\overline{\textsf{a}}}|v_{\bm{p}_3}|^2{+}\overline{\gamma}_{\textsf{a}}|v_{\bm{p}_4}|^2)\Bigr ]^{-1} .
\label{eq:res:tau:DD}
\end{align}
We emphasize that $1/\tau_{\textsf{a}}{\leqslant} 0$. 
If one 
neglects 
the single-particle decay rate $\bar{\gamma}_\textsf{a}|v_q|^2$, cf. Eq. \eqref{eq:G:SCBA}, in Eq.  \eqref{eq:res:tau:DD}, the function 
\begin{equation}
\Delta_{\gamma}(\bm{p}_1,\bm{p}_2,\bm{p_3},\bm{p_4})  \, \stackrel{\bar\gamma\to0}{\longrightarrow} \, \delta(\xi_{\bm{p}_1}{-}\xi_{\bm{p}_2}{-}\xi_{\bm{p}_3}{-}\xi_{\bm{p}_4}) .   
\end{equation}
Thus the result \eqref{eq:res:tau} resembles Fermi's Golden rule type expression. Its physical origin can be explained as follows. 
The diagrams from Fig. \ref{Figure:SE:diagrams1}, which give a non-zero contribution to $1/\tau_{\textsf{a}}$, can be expressed in terms of the contribution, $\Sigma^{R,(2)}_{\bm{p},\textsf{a}}(\varepsilon)$, of the second order in $\bar{\gamma}_\textsf{a}$ to the single particle retarded self-energy  (see Fig. \ref{Figure:SE:FGR}) as follows 
\begin{equation}
\frac{1}{\tau_{\textsf{a}}}{\simeq} - \frac{2 \int_{\bm{p_1}} [U^\dag_{\bm{p_1}}\hat\gamma_{\textsf{a}} U_{\bm{p_1}}]_{\textsf{aa}}|v_{\bm{p_1}}|^{-2}
\Im \Sigma^{R,(2)}_{\bm{p_1},\textsf{a}}(s_\textsf{a} \xi_{\bm{p_1}})}{\int_{\bm{k}} [U^\dag_{\bm{k}}\hat\gamma_{\textsf{a}} U_{\bm{k}}]_{\textsf{aa}} |v_{\bm{k}}|^{-2}}  .
\label{eq:rel:ImIn}
\end{equation}
Figure~\ref{Figure:SE:FGR} represents the process in which a fermion with momentum $\bm{p}_1$ in the $\textsf{u}$-band loses its energy by 
decaying into a three-particle state with two fermions with momenta $\bm{p_2}$ and $\bm{p_4}$ in $\textsf{u}$-band and one hole with momentum $\bm{p_3}$ in the $\textsf{d}$-band (similar process occurs for a hole in the $\textsf{d}$-band). 
Thus the number of particles in the $\textsf{u}$-band is increased. In semiconductors, such process is known as \textit{impact ionization} \cite{Keldysh1960}.  

{
The Fermi's Golden rule type of Eq.~\eqref{eq:res:tau} (with $\Delta_\gamma$ substituted by Dirac delta function) suggests 
that the rate of impact ionization process has a threshold in $\Im \Sigma^{R,(2)}_{\bm{p_1},\textsf{a}}(s_\textsf{a} \xi_{\bm{p_1}})$ for the momentum $\bm{p_1}$, see Eq. \eqref{eq:rel:ImIn}, due to the need to simultaneously satisfy 
the momentum and energy conservation laws
expressed by the Dirac delta functions \cite{Keldysh1960}. For example, for the model of Ref.~\cite{Tonielli2020} with $\xi_p{=}p^2{+}m^2$ the threshold is given as $|p_1|{\geqslant}\sqrt{3} m$. We note that the diffuson's self-energy $1/\tau_{\textsf{a}}$, Eq. \eqref{eq:res:tau}, involves $\Im \Sigma^{R,(2)}_{\bm{p_1},\textsf{a}}(s_\textsf{a} \xi_{\bm{p_1}})$ integrated over momentum $\bm{p}_1$, Eq. \eqref{eq:rel:ImIn}, such that the threshold would not be crucial for $1/\tau_{\textsf{a}}$.
In fact, 
due to broadening of the Green's function in the self-consistent Born approximation, cf. Eq. \eqref{eq:G:SCBA}, 
Eq.~\eqref{eq:res:tau} involves 
the Lorentzian-type expression $\Delta_\gamma$, see Eq. \eqref{eq:res:tau:DD}, that smears the hard threshold. 
}

In GKSL 
Eq.~\eqref{eq:GKSL}, 
the process of impact ionization appears because one of the fermion operators involved in $L_j$ is not an eigen-operator, but a linear combination of such. This leads to additional ``interference'' or mixed contributions when terms like $L^\dagger_j L_j$ are considered. 
Replacing the fermions
in our model with 
bosons results in the different sign for the second process in  Fig.~\ref{Figure:SE:FGR}. Consequently, $1/\tau_{\textsf{a}}$
in Eq.~\eqref{eq:res:tau} becomes positive.
Also the relative sign between the two terms in brackets in Eq.~\eqref{eq:res:tau} 
changes. Additionally, the last two diagrams in Fig.~\ref{Figure:SE:diagrams1} no longer cancel each other.

\begin{figure}[b]
\centerline{\includegraphics[width=1.04\columnwidth]{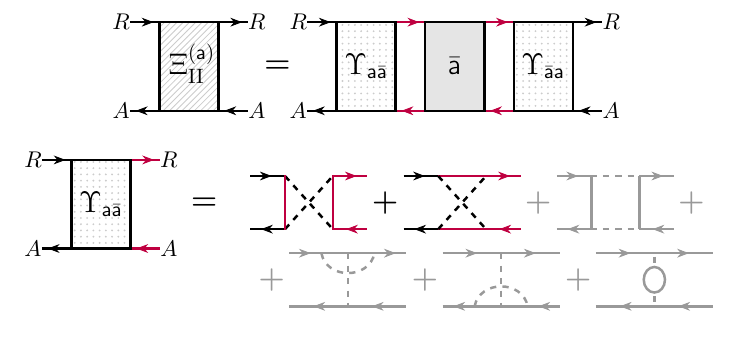}} 
\caption{The contribution to the self-energy $\Xi^{(\textsf{a})}_{\rm II}$, which corresponds to the dissipative drag, and which is obtained from the insertion of the $\bar{\textsf{a}}$-diffuson into the $\textsf{a}$-diffuson's ladder. Red lines denote the Green's function from the other ($\bar{\textsf{a}}$) band. Diagrams $\Upsilon_{\textsf{a}\bar{\textsf{a}}}$ are used to connect diffusons in different bands. Diagrams that could have made a contribution, but turned out to be equal to zero, are shown in pale color.}
\label{Figure:SE:diagrams2}
\end{figure}

\subsection{Dissipative drag}

The next type of the diffuson self-energy of the second order in $\bar{\gamma}_\textsf{a}$ allows combining diffusons from the different bands into a single ladder (see Fig.~\ref{Figure:SE:diagrams2}). Denoting the sum of diagrams in Fig.~\ref{Figure:SE:diagrams2} as $\Upsilon_{\textsf{a}\bar{\textsf{a}}}(\bm{Q},\Omega)$ we find that 
$\Upsilon_{\textsf{a}\bar{\textsf{a}}}(0,0){=}(\bar{\gamma}_{\bar{\textsf{a}}}/\bar{\gamma}_{\textsf{a}}) \Xi^{(\bar{\textsf{a}})}_{\rm I}(0,0)$ (see Appendix~\ref{App:InsertionDiffusons}). Summing the diffuson ladder with interchanged $\textsf{a}-$ and $\bar{\textsf{a}}-$diffusons
{
\begin{gather}
D_{\textsf{a}}^{(0)}{+}D_{\textsf{a}}^{(0)}\Upsilon_{\textsf{a}\bar{\textsf{a}}}D_{\bar{\textsf{a}}}^{(0)}\Upsilon_{\bar{\textsf{a}}\textsf{a}}D_{\textsf{a}}^{(0)}{+}D_{\textsf{a}}^{(0)}(\Upsilon_{\textsf{a}\bar{\textsf{a}}}D_{\bar{\textsf{a}}}^{(0)}\Upsilon_{\bar{\textsf{a}}\textsf{a}}D_{\textsf{a}}^{(0)})^2
{+}\dots \notag \\
= D_{\textsf{a}}^{(0)}/[1-\Upsilon_{\textsf{a}\bar{\textsf{a}}}D_{\bar{\textsf{a}}}^{(0)}\Upsilon_{\bar{\textsf{a}}\textsf{a}}D_{\textsf{a}}^{(0)}],
\end{gather}
}we find another contribution to the $\textsf{a}-$diffuson self-energy, $\Xi^{(\textsf{a})}_{\rm II}{\simeq}{-}[\tau_{\textsf{a}}\tau_{\bar{\textsf{a}}}(D^{(\bar{\textsf{a}})}_{jl}Q_jQ_l{-}i\Omega)]^{-1}$ (see Appendix~\ref{App:InsertionDiffusons}). The diagrams in Fig.~\ref{Figure:SE:diagrams2} resemble corrections to a Coulomb drag between two species of fermions (in that case, the dashed line denotes Coulomb interaction) \cite{Narozhny2016}, with the important difference that our dissipative interaction does not conserve the isospin associated with the $\textsf{u/d}$ band index.  
Therefore, we termed the discussed above contribution $\Xi^{(\textsf{a})}_{\rm II}$
as a \textit{dissipative drag} contribution. 

Although $\Xi^{(\textsf{a})}_{\rm II}$ is formally of the forth 
 order in $\bar{\gamma}_\textsf{a}$, since it contains the diffusive pole in the denominator, effectively, it acts as if it is of the second 
 order in $\bar{\gamma}_\textsf{a}$. 
To demonstrate it we combine the above results for the self-energies due to the impact ionization process and dissipative drag into the following system of equations for the diffusons in $\textsf{u/d}$-bands,
\begin{gather}
\Bigl (D^{(\textsf{a})}_{jl}Q_jQ_l{-}i\Omega{-}\frac{1}{|\tau_{\textsf{a}}|}\Bigr ) D_{\textsf{a}}(\bm{Q},\Omega) {+}
\frac{1}{|\tau_{\bar{\textsf{a}}}|}D_{\bar{\textsf{a}}}(\bm{Q},\Omega) \notag \\{=}2\bar{\gamma}_{\textsf{a}}^2\left [\int_{\bm{k}}\frac{[U_{\bm{k}}^{\dagger}\hat{\gamma}^{(\textsf{a})}U_{\bm{k}}]_{\textsf{aa}}}{\left|v_{\bm{k}}\right|^{2}}\right ]^{-1}  .
\label{eq:res:DD}
\end{gather}
As one can see, both the impact ionization process and dissipative drag provide the contribution of the same order, 
$1/|\tau_{\textsf{a}}|$ and $1/|\tau_{\bar{\textsf{a}}}|$, respectively.

\subsection{Non-radiative recombination}

In previous calculations of the diffuson self-energy, we used the Green's functions \eqref{eq:G:SCBA}, corresponding to the DS.
There is another type of self-energy corrections related to considering fermion density deviations $\delta n_{\textsf{a}}$ from the DS.
Nonzero $\delta n_{\textsf{a}}$ modifies the single particle self-energy, $\Sigma_{\bm{p}}(\varepsilon)$, already at the level of the Hartree-Fock diagrams (see Fig.~\ref{Fig:SCBA}). Using such modified (by $\delta n_{\textsf{a}}$) single particle Green's function in the self-energy for calculation of the diffusive ladder, we find that the diffusion pole is modified by appearance of the self-energy $\Xi^{(\textsf{a})}_{\rm III}$. To the lowest order in density deviations and $\bar{\gamma}_\textsf{a}$, we find 
\begin{equation}
\Xi^{(\textsf{a})}_{\rm III}(0,0){=}{-}\beta s_{\textsf{a}}\delta n_{\bar{\textsf{a}}}
\label{eq:Xi:III}
\end{equation}
 with
$\beta{=}\beta_{\textsf{u}}{+}\beta_{\textsf{d}}$, where (see Appendix \ref{App:Recombination})
\begin{equation}
\beta_{\textsf{a}}=\frac{2\overline{\gamma}_{\textsf{a}} \int_{\bm{p}}[U_{\bm{p}}^{\dagger} \hat{\gamma}_{\textsf{u}}U_{\bm{p}}]_{\textsf{aa}}[U_{\bm{p}}^{\dagger} \hat{\gamma}_{\textsf{d}}U_{\bm{p}}]_{\textsf{aa}}|v_{\bm{p}}|^{-2}}{\int_{\bm{k}}[U_{\bm{k}}^{\dagger}\hat{\gamma}_{\textsf{u}}U_{\bm{k}}]_{\textsf{uu}}|v_{\bm{k}}|^{-2}\int_{\bm{q}}[U_{\bm{q}}^{\dagger}\hat{\gamma}_{\textsf{d}}U_{\bm{q}}]_{\textsf{dd}}|v_{\bm{q}}|^{-2}} .
\label{eq:res:beta}
\end{equation}
We note that the self-energy $\Xi^{(\textsf{a})}_{\rm III}$ of the diffuson in the $\textsf{a}$-band is proportional to the density deviation in the other $\bar{\textsf{a}}-$band. Such a process is known
in semiconductors as 
\textit{band-to-band non-radiative recombination}~\cite{Abakumov-book}.  
There is no contribution to $\Xi^{(\textsf{a})}$ proportional to $\delta n_{\textsf{a}}$ in the first order in $\bar{\gamma}_\textsf{a}$. Such contribution arises in the second order in $\bar{\gamma}_\textsf{a}$ 
only 
and corresponds to the Auger recombination~\cite{Abakumov-book}, the inverse process to the impact ionization
~\footnote{Strictly speaking, the term due to Auger recombination should be proportional to $n_u n_d$. However, since we are working in the regime with $n_d\simeq n+\delta n_d$ and $n_u\simeq \delta n_u$, we find $n_u n_d\simeq n \delta n_u$.}.

\section{FKPP-equation\label{Sec:FKPP}}

Combining Eq. \eqref{eq:Xi:III} for the non-radiative recombination with Eq.~\eqref{eq:res:DD} for the diffuson in the presence of the impact ionization and dissipative drag,  
we can cast the above results as 
equations for the density deviation $\delta n_{\textsf{a}}(\bm{x},t)$ from the 
DS (
with $n_{\textsf{d}}{=}n$ and $n_{\textsf{u}}{=}0$): 
\begin{equation}
\begin{split}
\Bigl (\partial_t{-}D^{(\textsf{u})}_{jl}\nabla_j\nabla_l{-}\frac{1}{|\tau_{\textsf{u}}|}-\beta \delta n_{{\textsf{d}}}\Bigr ) \delta n_{\textsf{u}} {+}
\frac{\delta n_{\textsf{d}}}{|\tau_{{\textsf{d}}}|} {=}0  ,
\\
\Bigl (\partial_t{-}D^{(\textsf{d})}_{jl}\nabla_j\nabla_l{-}\frac{1}{|\tau_{\textsf{d}}|}+\beta \delta n_{{\textsf{u}}}\Bigr ) \delta n_{\textsf{d}} {+}
\frac{\delta n_{\textsf{u}}}{|\tau_{{\textsf{u}}}|} {=}0 .
\end{split}
\label{eq:res:FKPP}
\end{equation}

We note that the diagonal components of the inverse of the matrix operator in the right-hand side of Eq. \eqref{eq:res:FKPP} reproduces the diffuson with the self-energy $\Xi^{(\textsf{a})}$ discussed above. It is convenient to introduce the total density deviation $\delta n{=}\delta n_{\textsf{u}}{+}\delta n_{\textsf{d}}$ and the imbalance $\delta m{=}\delta n_{\textsf{u}}{-}\delta n_{\textsf{d}}$.
Then Eq. \eqref{eq:res:FKPP} can be rewritten as 
\begin{gather}
    \Bigl (\partial_t{-}D^{(+)}_{jl}\nabla_j\nabla_l{-}\frac{1}{\tau_+}\Bigr )\delta m {+}\frac{\beta}{2}(\delta m)^2 {=} D^{(-)}_{jl}\nabla_j\nabla_l \delta n {+} \frac{\delta n}{\tau_-} \notag \\
    {+} \frac{\beta}{2}(\delta n)^2 ,\quad
    \Bigl (\partial_t{-}D^{(+)}_{jl}\nabla_j\nabla_l\Bigr )\delta n = D^{(-)}_{jl}\nabla_j\nabla_l \delta m ,  
    \label{eq:res:FKPP:2}
\end{gather}
where $D^{(\pm)}_{jl}{=}(D^{(\textsf{u})}_{jl}{\pm}D^{(\textsf{d})}_{jl})/2$ and $\tau_\pm^{-1}{=}|\tau_{\textsf{u}}|^{-1}{\pm}|\tau_{\textsf{d}}|^{-1}$. According to Eq. \eqref{eq:res:FKPP:2}, the total density deviation is governed by a pure diffusive equation that is the consequence of the conservation of the total number of fermions. At half-filling, one finds $\int_{\bm{x}} \delta n(\bm{x},t){=}0$. Contrary to $\delta n$, the imbalance is governed by 
the FKPP equation describing typically
a reaction-diffusion type dynamics \cite{Fisher1937,Kolmogorov1937,FKPP1988,FKPP2,FKPP3,Aleiner2016,Zhou2023}.

Let us discuss possible homogeneous stationary solutions of Eq.~\eqref{eq:res:FKPP:2} at half-filling. In this case, the total particle density vanishes, $\delta n{=}0$. For the imbalance, we find an unstable solution, $\delta m{=}0$, corresponding to the engineered DS.
There is also a stable solution $\delta m_{*} {=} 2/(\beta \tau_+)$, appearing at the first order in $\bar{\gamma}_\textsf{a}$, such that the density $|\delta n_{\textsf{a}}|{=}\delta m_{*}/2{\ll} n$. The imbalance at the stable stationary state can be estimated from above as follows.

With the inequality for an arbitrary unitary matrix $U_{\bm{q}}$, 
\begin{equation}\label{eq:min_max_U}
   \min\{\hat{\gamma}_{\textsf{a}}\} \leqslant \left[U^{\dag}_{\bm{p}_1}\hat{\gamma}_{\textsf{a}} U_{\bm{p}_1}\right]_{\textsf{a}\textsf{a}} \leqslant   
   \max\{\hat{\gamma}_{\textsf{a}}\} ,
\end{equation}
and the inequality
\begin{equation}
|A_{\textsf{ud}}|^2+|A_{\textsf{du}}|^2= \tr A^\dag A - |A_{\textsf{uu}}|^2-|A_{\textsf{dd}}|^2
\leqslant  \tr A^\dag A, 
\end{equation}
we find 
\begin{align}
 \delta m_* & \leqslant \frac{\max\{\hat{\gamma}_{\textsf{u}}\}\max\{\hat{\gamma}_{\textsf{d}}\}[\max^2\{\hat{\gamma}_{\textsf{u}}\}+\max^2\{\hat{\gamma}_{\textsf{d}}\}]} {\min\{\hat{\gamma}_{\textsf{u}}\}\min\{\hat{\gamma}_{\textsf{d}}\}[\min\{\hat{\gamma}_{\textsf{u}}\}+\min\{\hat{\gamma}_{\textsf{d}}\}]}\notag \\
 & \times \int_{\bm{p}_i}
\delta(\bm{p}_1{-}\bm{p}_2{+}\bm{p}_3{-}\bm{p}_4) \delta(\xi_{\bm{p}_1}{-}\xi_{\bm{p}_2}{-}\xi_{\bm{p}_3}{-}\xi_{\bm{p}_4})\notag \\
 & \times \tr \bigl [v_{\bm{p_4}}^* U^\dag_{\bm{p_2}}-v_{\bm{p_2}}^* U^\dag_{\bm{p_4}}\bigr ]
 \bigl [v_{\bm{p_4}}U_{\bm{p_2}}-v_{\bm{p_2}} U_{\bm{p_4}}\bigr ] .
    \end{align}
Here the trace in the last line can be written in a manifestly positive form $\sum_{\textsf{a},\alpha}
\bigl |[v_{\bm{p_4}}U_{\bm{p_2}}{-}v_{\bm{p_2}} U_{\bm{p_4}}]_{\alpha,\textsf{a}}\bigr |^2 $, which in turn can be bounded as follows 
\begin{gather}
 \sum_{\textsf{a},\alpha}
\bigl |[v_{\bm{p_4}}U_{\bm{p_2}}{-}v_{\bm{p_2}} U_{\bm{p_4}}]_{\alpha,\textsf{a}}\bigr |^2 
{\leqslant}\sum_{\textsf{a},\alpha}
\Bigl ( |v_{\bm{p_4}} [U_{\bm{p_2}}]_{\alpha,\textsf{a}}|\notag \\ {+}|v_{\bm{p_2}}[U_{\bm{p_4}}]_{\alpha,\textsf{a}}| \Bigr )
{\leqslant}4(|v_{\bm{p_4}}|^2{+}|v_{\bm{p_2}}|^2) .
\end{gather}
From this, we finally derive the estimate
\begin{gather}
\delta m_{*}  \leqslant \frac{4\pi}{n} 
\frac{\max\{\hat{\gamma}_{\textsf{u}}\}\max\{\hat{\gamma}_{\textsf{d}}\}[\max^2\{\hat{\gamma}_{\textsf{u}}\}+\max^2\{\hat{\gamma}_{\textsf{d}}\}]} {\min\{\hat{\gamma}_{\textsf{u}}\}\min\{\hat{\gamma}_{\textsf{d}}\}[\min\{\hat{\gamma}_{\textsf{u}}\}+\min\{\hat{\gamma}_{\textsf{d}}\}]}
\notag \\
 \times 
\int_{\bm{p}_i}
 \delta(\xi_{\bm{p}_2{-}\bm{p}_3{+}\bm{p}_4}{-}\xi_{\bm{p}_2}{-}\xi_{\bm{p}_3}{-}\xi_{\bm{p}_4}) (|v_{\bm{p_4}}|+|v_{\bm{p_2}}|)^2 .
\label{eq:res:estimate}
\end{gather}

\begin{figure*}[t!]
\centerline{\includegraphics[width=1.0\textwidth]{FKPPNew5.png}}
\caption{ Numerical solution of the FKPP equations \eqref{eq:res:FKPP} in $d=1$ on a finite interval $x\in ({-}L,L)$ with $L=100$ and periodic boundary conditions, for several values of $D^{(-)}/D^{(+)}$ and $\tau_+/\tau_-$. Here the distance $x$ is measured in units of $\sqrt{\tau_+ D^{(+)}}$, the time $t$ is in units of $\tau_+$, and the density deviations are in units of $2/\beta\tau_+$. In all cases, the initial condition is $\delta n_{\textsf{u/d}}(x,0)= \pm 0.1 \exp\left\{-2L\sin^2(\pi (x\mp L/4)/2L )\right\}$. (a-d): The density deviations for the following parameters $\{D^{(-)}/D^{(+)}, \;\tau_+ /\tau_-\} = \{0,\;0.99\},\;\{0.99,\;0.99\},\; \{0.99,\;-0.99\},\; \{0.99,\;0\} $ (from left to right), with each row corresponding to one of the densities, i.e. $\delta n_{\textsf{u}}$, $|\delta n_{\textsf{d}}|$, $\delta n{=}\delta n_{\textsf{u}}{+}\delta n_{\textsf{d}}$, or $\delta m{=}\delta n_{\textsf{u}}{-}\delta n_{\textsf{d}}$ (from top to bottom).}
\label{fig:FKPP2}
\end{figure*}


The spatial and temporal dynamics of $\delta n_{\textsf{u}}$ and $\delta n_{\textsf{d}}$, as governed by Eq.~\eqref{eq:res:FKPP:2}, closely resemble those of the standard FKPP equation \cite{FKPP-book,Grindrod-book}: at early times, the densities spread diffusively, and at longer timescales, a ballistic front emerges (see Fig.~\ref{fig:FKPP2}). However, unlike the one-component FKPP equation---which, after suitable rescaling, becomes entirely universal with no adjustable parameters---the two-component system described by Eq.~\eqref{eq:res:FKPP:2} depends crucially on the relation between the diffusion coefficients, $D^{(-)}/D^{(+)}$, and the impact ionization times, $\tau_{+}/\tau_{-}$. To illustrate how these parameters influence the dynamics, we solved Eq.~\eqref{eq:res:FKPP:2} numerically on a one-dimensional interval $(-L, L)$ with periodic boundary conditions. The initial configuration consisted of two spatially separated peaks in the local density deviations from half-filling, arranged so that the total integrated deviation is zero. Specifically, we placed an excess of particles in the upper band near $x {=} L/4$, while an equivalent number of holes in the lower band was localized near $x {=} {-}L/4$.

We first consider the case of identical diffusion constants $D^{({\textsf{u}})}{=}D^{({\textsf{d}})}$ (i.e. $D^{(-)}/D^{(+)} {=} 0$), but with very different impact ionization rates, $ 1/|\tau_{\textsf{u}}| {\gg} 1/|\tau_{\textsf{d}}|$. Under this condition, the density in the lower band decays very rapidly in regions where upper-band particles are present (and thus $\delta n_{\textsf{d}}$ quickly grows in magnitude). Fig.~\ref{fig:FKPP2}(a) confirms this fast ``inflation'' of a second peak in $\delta n_{\textsf{d}}$ at $x {=} L/4$, which develops much more quickly than the initial peak at $x {=} {-}L/4$. After this rapid initial phase, both peaks evolve into ballistically propagating fronts traveling at the same velocity. Next, we include strong anisotropy in the diffusion coefficients, $D^{(-)}/D^{(+)} {\sim} 1$, which strongly affects the ``inflation" stage (Fig.~\ref{fig:FKPP2}(b)). In this regime, the lower-band density fluctuation barely diffuses on its own (as evidenced by the negligible spreading of the initial peak at $x {=} {-}L/4$ in the second row in Fig.~\ref{fig:FKPP2}(b)). Once the local densities in both bands become comparable, however, the dissipative drag again drives a ballistic front. Because down-band particles are slow, the total density deviation $\delta n = \delta n_{\textsf{u}} + \delta n_{\textsf{u}}$ at the front is positive (the front contains more up-band particles than down-band holes), leaving behind a negative deviation at the original position of the $\delta n_{\textsf{d}}$ peak at $x{=}{-}L/4$. When $\tau_{+}/\tau_{-}$ is close to ${-}1$ (but still with $D^{(-)}/D^{(+)}{\sim} 1$), the spatial dynamics of the down-band holes becomes very slow, leading to a reduced front velocity compared to the previous scenario (Fig.~\ref{fig:FKPP2}(c)). Finally, in the case of identical impact ionization times, $|\tau_{\textsf{u}}| {=} |\tau_{\textsf{d}}|$, the late-time evolution is fairly symmetric across the two bands---even if their diffusion constants differ significantly---resulting in nearly identical fronts in both $\delta n_{\textsf{u}}$ and $\delta n_{\textsf{d}}$, as shown in Fig.~\ref{fig:FKPP2}(d).

We also note that the two-component FKPP equation \eqref{eq:res:FKPP} shares certain similarities with other well-studied reaction-diffusion universality classes, such as two-component Fitzhugh–Nagumo systems \cite{FitzHugh1955,Nagumo1962}. However, to the best of our knowledge, the complete classification of all possible dynamical regimes and traveling wave solutions of Eq.~\eqref{eq:res:FKPP} remains an unsolved mathematical problem. It would be interesting to see how the late-time and long-distance dynamics predicted by Eq.~\eqref{eq:res:FKPP} emerges in a direct numerical solution of the GKSL master equation \eqref{eq:GKSL} for our model.


\section{Discussions\label{Sec:Discussions}}

The most significant outcome of Eq.~\eqref{eq:res:FKPP:2} is the instability of the 
engineered DS with a completely empty upper band, $\delta m{=}\delta n{=}0$. This 
instability is caused by the negative rate $1/\tau_{\textsf{a}}$, which has a clear physical origin in the process of impact ionization, generating fermions in the $\textsf{u}-$band and holes in the $\textsf{d}-$band. The FKPP 
Eq.~\eqref{eq:res:FKPP:2} predicts another stable steady state with non-zero $\delta n_{\textsf{a}}{=}{\pm}\delta m_*/2$,  suggesting that the models under consideration have a dark space rather than a unique 
DS. It would be interesting to test this result against recent proofs of the uniqueness of a 
DS for GSKL equation on a lattice \cite{Kawabata2024,Yoshida2024}. We also note that various sufficient conditions for the emergence of multiple dark states have been discussed in the literature (see, for example, \cite{Buca2012,Albert2014}), often emphasizing the role of strong symmetries and conserved quantities. While a strong $U(1)$ symmetry is indeed a fundamental feature of our model, we stress that our instability mechanism (as described in Sec.~\ref{SubSec:ImpIon} and illustrated in Fig.~\ref{Figure:SE:FGR}) has a more subtle origin. Specifically, it arises from the interference or ``mixed" processes in Eq.~\eqref{eq:GKSL}, involving terms of the form $L^\dagger L$. Since our results are derived in regime of weak dissipation, $\bar{\gamma}_{\textsf{a}}{\ll}1$, we cannot exclude the possibility that the 
engineered DS is stabilized with increasing $\bar{\gamma}_{\textsf{a}}$, allowing for a potential phase transition between two different steady states.

In the weak dissipation regime, $\bar{\gamma}_{\textsf{a}}{\ll}1$, one can avoid the instability of the dark state if $1/\tau_{\textsf{a}}$ vanishes identically. This requires complete destructive interference of the two processes shown in Fig.~\ref{Figure:SE:FGR}. The off-shell condition for this to happen is very restrictive: $U_{\bm{p}}{=}v_p \mathcal{U}$ and $|v_p|{=}1$, where $\mathcal{U}$ is an arbitrary constant unitary matrix. As an example of on-shell complete destructive interference, we can mention a $d{=}1$ version of the model of Ref.~\cite{Tonielli2020} with $\gamma_{\textsf{a},\alpha}{=}\gamma$. In this case, one has to compute corrections of the third order in $\bar{\gamma}_{\textsf{a}}$ to the diffuson's self-energy \cite{elsewhere}. However, already for $\gamma_{\textsf{a},1}{\neq}\gamma_{\textsf{a},2}$ this  complete destructive interference disappears, and there is a nonzero $1/\tau_{\textsf{a}}$ in the second order in $\bar{\gamma}_{\textsf{a}}$. Additionally, a finite magnitude of $1/\tau_{\textsf{a}}$ can be obtained in models with a momentum-independent rotation matrix $U_{\bm{p}}$ and arbitrary momentum-dependent $v_p$.

Another possible way to avoid the instability inherent in Eq.~\eqref{eq:res:tau} is the  
limiting case where the energy conservation condition 
cannot be fulfilled (
e.g., for $\xi_p{=}\textrm{const}$). In this scenario, one must account for the dissipation-induced broadening of the 
$\delta$-function in Eq.~\eqref{eq:res:tau} beyond Fermi's Golden rule (see 
Eq.~\eqref{eq:AppA:Xi1}). This  results in 
$1/\tau_{\textsf{a}}{\propto}\bar{\gamma}_\textsf{a}^3$, so that further analysis of other $\mathcal{O}(\bar{\gamma}_\textsf{a}^3)$ contributions is required, which we did not attempt here. In the other extreme case without any Hamiltonian evolution  ($\xi_p{\equiv}0$), the system is always at strong coupling. This is evident by simply rescaling the time variable in Eq.~\eqref{eq:GKSL} as $t{\rightarrow} t/\operatorname{min}\{\bar{\gamma}_{\textsf{a}}\}$, eliminating a small parameter. This is inconsistent with our initial assumption that the deviation from the 
DS parametrically small. Thus, 
more involved analysis is needed than that presented above.

It is also important to emphasize that our results obtained in the continuous limit may not necessarily hold 
for small finite-size systems. The point lies in the discreteness of the spectrum $\xi_{\bm{p}}$ in a system of a finite size $L$. 
As well known \cite{Sivan1994,Blanter1996,Altshuler1997,Mirlin1997,Silvestrov1997,Silvestrov2001,Auerbach2011,Gornyi2016s,Gornyi2017s,Micklitz2022}, the computation of decay and broadening for discrete levels is a complicated problem. Our computations of rates for the impact ionization process and dissipative drag were based essentially on the Fermi's Golden rule approximation. They are both related to a decay of a  fermion, e.g. from the $\textsf{u}$-band, into the three particle state, see Fig.~\ref{Figure:SE:FGR}. The corresponding three-particle level spacing can be estimated as 
\begin{equation}
\frac{1}{\Delta_3(\xi_{\bm{p}})} {=} L^{3d}\int_{\bm{p_j}}
\delta\left (\xi_{\bm{p}} {-} \xi_{\bm{p_1}}{-}\xi_{\bm{p_2}}{-}\xi_{\bm{p_3}}\right ) .
\label{eq:Delta3}
\end{equation}
In order for the process to be efficient and its rate to be described by the Fermi's Golden rule, the magnitude of the energy broadening in Eq. \eqref{eq:rel:ImIn}
has to be larger than 
the three particle level spacing \cite{Altshuler1997}, 
\begin{equation}
|\Im \Sigma^{R,(2)}_{\bm{p},\textsf{u}}(\xi_{\bm{p}})|\gg \Delta_3(\xi_{\bm{p}}) .
\label{eq:ImS:cond}
\end{equation}
The above inequality limits the system sizes for which Fermi's Golden rule is applicable from below, $L{\gg}\max L_{\xi_{\bm{p}}}$. For a gapped spectrum $\xi_{\bm{p}}$, both $\Im \Sigma^{R,(2)}_{\bm{p},\textsf{u}}(\xi_{\bm{p}})$ and $1/\Delta_3(\xi_{\bm{p}})$ vanish provided that $\xi_{\bm{p}}$ is below the threshold value, and thus, the length $L_{\xi_{\bm{p}}}$ tends to infinity for such $\xi_{\bm{p}}$. As we discussed in Sec.~\ref{SubSec:ImpIon} the hard thresholds related with the Dirac delta functions due to energy conservation are relaxed in virtue of the Green's function broadening within the self-consistent Born approximation. Still it makes the length $L_{\xi_{\bm{p}}}$ finite but large as $\bar{\gamma}_{\textsf{a}}{\to} 0$. Therefore, to test the applicability of our predictions to a numerical simulation or physical experiment one has to carefully estimate the system sizes needed.


In order to provide a more concrete criterion for how large the system size has to be, we apply the condition from Eq.~\eqref{eq:ImS:cond} to the impact ionization rates given in Eq.~\eqref{eq:rel:ImIn}, which, in combination with the bound Eq.~\eqref{eq:min_max_U}, leads to 
\begin{equation}\label{eq:tau_estimate}
    |1/\tau_{\textsf{a}}|\gg \frac{2\min\{\hat{\gamma}_{\textsf{a}}\} }{\max\{\hat{\gamma}_{\textsf{a}}\} \int_{\bm{k}} |v_{\bm{k}}|^{-2} }\int_{{\bm{p}}}\frac{\Delta_3(\xi_p)}{|v_{\bm{p}}|^2}\;
\end{equation}
for any choice of the unitary matrices $U_{\bm{p}}$. For example, for the model of a topological insulator from Ref.~\cite{Tonielli2020}, in which $\xi_{\bm{p}}{=}p^2{+}m^2$, $v_{\bm{p}}{=}\sqrt{\xi_{\bm{p}}}$, 
and $\bar{\gamma}_{\textsf{a}}{\equiv}\bar{\gamma}$, one finds 
\begin{equation}
 \frac{1}{\Delta_3}  {=} \frac{L^{3d} S_{3d}}{2(2\pi)^{3d}}\left [(\xi_{\bm{p}}{-}3m^2)^{\frac{3d{-}2}{2}}\Theta(\xi_{\bm{p}}{-}3m^2) {+} a_d \bar{\gamma} n^{3{-}\frac{2}{d}}\right ],
 \label{eq:dd3:S}
\end{equation}
where $\Theta(x)$ denotes the Heaviside step function, $a_d{=}2(2\pi)^{3d{-}2}(3d/S_{3d})^{1{-}2/(3d)}/(3d{-}2)$, and $S_{3d}=2\pi^{3d/2}/\Gamma(3d/2)$ stands for the area of the unit sphere in $3d$ dimensional space. The last term in Eq. \eqref{eq:dd3:S} originates from the broadening of the Dirac delta function in Eq. \eqref{eq:Delta3} due to imaginary part of the Green's function within the self-consistent Born approximation, cf. Eq. \eqref{eq:G:SCBA}. Now let's focus on the dimensions $d{=}1$ and $d{=}2$.
After performing the integration on the r.h.s. of Eq.~\eqref{eq:tau_estimate}, 
and using the continuous-limit result $|1/\tau_{\textsf{u}}|{=}b_d \bar{\gamma}^2 m^2$ with $b_1{=}\pi/2$ and $b_2{=}8/3$ for $\bar{\gamma} n/m^d{\gg}1$  (see Appendix~\ref{App:Estimation}), we find the following estimate for the system size, 
\begin{equation}
    n L^d\gg \frac{c_d}{\bar{\gamma}} \left (\frac{n}{m^d}\right)^{2/(3d)} , \qquad \bar{\gamma} n/m^d \gg 1.
\end{equation}
Here we introduced the numerical constant $c_d{=}(2\pi)^d[S_{3d} a_d b_d/4]^{-1/3}$. 
We note that the above inequality is satisfied already at the length $L{\sim}\sqrt{\tau_{+}D^{(+)}}$.

\color{black}

It should also be noted that all parameters in the FKPP Eq.~\eqref{eq:res:FKPP} will acquire higher-order fluctuation-induced corrections, which involve insertions of additional diffusons in the internal loop integrals. In sufficiently low spatial dimensions, these corrections could lead to scale-dependent renormalizations and affect stability of steady states. 
As known \cite{Cardy1996,cardy1998field,Canet2004,Canet_2006}, 
such fluctuation effects can induce absorbing phase transitions in classical stochastic reaction-diffusion models described by FKPP-like equations at the mean-field level. This provides further support for a possible phase transition between two steady states in our problem. We leave the investigation of 
fluctuation corrections for future work.

Finally, we mention that the instability of the dark state can be affected by the presence of interaction between fermions and by a quenched disorder in the Hamiltonian part of the GSKL equation, Eq.~\eqref{eq:GKSL}. In particular, dissipation-induced diffusion in our model, combined with additional fermion interactions, is likely to produce a strong scale-dependent renormalization of the local density of states and conductivity, analogous to the well-known Altshuler-Aronov effect in disordered conductors \cite{ALTSHULER19851}. Formally, interactions introduce an additional quartic operator in the Keldysh field theory action, which, unlike the number-conserving dissipation examined here, does not couple the two branches of the Keldysh contour \cite{Thompson2023}. The methods developed in this work allow for a straightforward inclusion of this modification within perturbation theory.
However, in the second-order perturbation theory the dissipation-induced contribution to the impact ionization, $1/\tau_{\textsf{a}}$, will not interfere with the contribution from the electron-electron interaction \cite{Keldysh1960}. The reason is that diagrams in Fig. \ref{Figure:SE:diagrams1} that do not vanish due to causality in the time domain (all diagrams except the first two diagrams in the second line of Fig. \ref{Figure:SE:diagrams1}) involve a single Green's function from the other band. This implies that both dashed lines change the band index on one of their two sides. Since electron-electron interactions do not exhibit this property, one cannot replace one of dissipation-induced dashed lines with an interaction line. Therefore, we expect the dissipation-induced and interaction-induced contributions to the ionization rates to be simply additive (when both dissipation and interactions are weak). As a result, our conclusions should remain valid within at least some range of parameters in the interacting model.  
\color{black}


 \section{Summary\label{Sec:Summary}}
 
 To summarize, we studied particle-number-conserving dissipative dynamics  for a class of two-band fermionic models. Although the jump operators in these models are engineered to guarantee the dark state with the completely empty upper band, we found that this dark state is generically unstable, at least in a weak dissipation regime. We discovered that this instability originates from impact ionization, a physical process well known in semiconductors. We derived equations governing the evolution of densities in each band, resembling FKPP reaction-diffusion equations. These equations indicate a stable steady state with a nonzero population in the upper band. Even though the instability we found is very general, posing severe limitations on the possibility of stabilizing dark states with particle-conserving dissipation, we have outlined several fine-tuned situations that circumvent it. These conditions could be utilized in designing future implementations of dissipative protocols.

In the future, we aim to extend our research in two directions: (i) deriving the kinetic equation for the distribution function which is a $2{\times}2$ matrix in band space (essentially, a kind of semiconductor Bloch equation, see e.g. Ref.~\cite{Quade1994}), which will allow us to understand energy resolved interplay between impact ionization and non-radiative recombination and how particles in a stable steady state are distributed in the momentum space and (ii) examining the stability of the dark state under strong dissipation, fluctuation effects, as well as the presence of electron-electron interaction. Additionally, it would be beneficial to test our predictions using numerical simulations of lattice models.

\begin{acknowledgements}
We are grateful to A. Altland, S. Das, S. Diehl, A. Kamenev, N. O’Dea, and, especially, to M. Glazov, for useful discussions. 
We thank S. Diehl for sharing results of numerical simulations prior to publication.
We thank M. Goldstein for collaboration on the initial stage of this project. The research of I.S.B. and A.A.L. was supported by the Ministry of Science and Higher Education of the Russian Federation (project no. FFWR-2024-0017). The work of P.A.N. was supported in part by the US Department of Energy, Office of Basic Energy Sciences, Division of Materials Sciences and Engineering, under contract number DE-AC02-76SF00515. I.S.B. and A.A.L. acknowledge the hospitality during the 
``Nor-Amberd School in Theoretical Physics 2024'' where part of this work has been performed. I.S.B. and A.A.L. acknowledge personal support from the Foundation for the Advancement of Theoretical Physics and Mathematics ``BASIS''. 
\end{acknowledgements}

\onecolumngrid
\appendix

\section{Calculation of the self-energy. \label{App:SE}}

In this section, we calculate the self-energy contribution $\Xi^{(\textsf{a})}_{\rm I}$, which consists of the diagrams, depicted in Fig. \ref{Figure:SE:diagrams1}.

\begin{figure}[h!]
\centerline{\includegraphics[width=0.6\columnwidth]{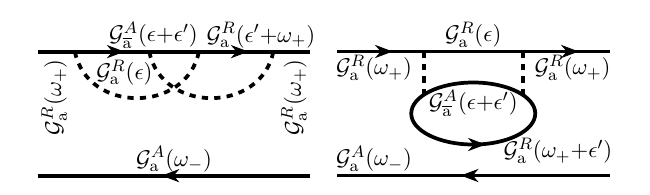}} 
\caption{Arrangement of indices on diagrams that make a real contribution to $\Xi^{(\textsf{a})}_{\rm I}(0,0)$. Complex conjugates are omitted. The arrangement of matrices $\overline{\mathcal{L}}$ and $\mathcal{L}$ on these diagrams is uniquely determined by $\textsf{a}{=}\textsf{u}$ or $\textsf{a}{=}\textsf{d}$.}
\label{Fig:SII:MainDiagrams}
\end{figure}

Let's look at diagrams that give a non-zero contribution (indicated in full color in Fig. \ref{Figure:SE:diagrams1}). Some arrangements of indices on the diagrams turn out to be zero. To select the right ones, we need to keep an eye on four things: (1) the structure of Keldysh dissipative action; (2) the matrix representation of the single-particle Green's functions, $\mathcal{G}_{\bm{p},\textsf{a}}(\varepsilon){=}\mathcal{G}^R_{\bm{p},\textsf{a}}(\varepsilon) \Lambda^{(+)}_\textsf{a} {+} \mathcal{G}^A_{\bm{p},\textsf{a}}(\varepsilon) \Lambda^{(-)}_\textsf{a}$ (see Eq.~\eqref{eq:Lambda}); (3) the definitions of the four matrices $\mathcal{L}^{(\textsf{a},\alpha)}$ (see Eq.~\eqref{eq:A:matrices:0}); (4) location of the poles of Green's functions when integrating over frequencies. After taking into account all the details, there will be only one possible arrangement of indices left on each diagram, see Fig.~\ref{Fig:SII:MainDiagrams}.

Remembering the definition \eqref{SI:Sigma}, we get
\begin{gather}
	\Sigma^{(\textsf{a})}_{\rm I}(\bm{Q},\Omega)=\int_{\bm{q},\omega,\bm{p},\varepsilon}v_{\bm{q}_{+}}v_{\bm{q}_{-}}^*\left[U_{\bm{p}_{+}}^{\dagger} \hat{\gamma}^{(\textsf{a})}U_{\bm{p}_{-}}\right]_{\textsf{aa}}\mathcal{G}_{\bm{q}_{+},\textsf{a}}^R(\omega_{+}) \mathcal{G}_{\bm{q}_{-},\textsf{a}}^A(\omega_{-}) \mathcal{G}_{\bm{p}_{+},\textsf{a}}^R(\varepsilon_{+})\int_{\epsilon, \bm{k},\epsilon', \bm{k}'}(2\pi)^{d+1}\delta(\omega-\varepsilon)\delta(\bm{q}-\bm{p})\notag \\
   \times \Big\{v_{\bm{k}'+\bm{q}_+}v_{\bm{k}}^*|v_{\bm{q}_+}|^2\left[U_{\bm{k}'+\bm{q}_+}^{\dagger} \hat{\gamma}^{(\textsf{a})} U_{\bm{k}+\bm{k}'}\right]_{\textsf{a}\overline{\textsf{a}}}\left[U_{\bm{k}+\bm{k}'}^{\dagger} \hat{\gamma}^{(\textsf{a})} U_{\bm{k}}\right]_{\overline{\textsf{a}}\textsf{a}}-|v_{\bm{q}_++\bm{k}'}|^2|v_{\bm{q}_+}|^2\left[U_{\bm{k}}^{\dagger} \hat{\gamma}^{(\textsf{a})} U_{\bm{k}+\bm{k}'}\right]_{\textsf{a}\overline{\textsf{a}}}\left[U_{\bm{k}+\bm{k}'}^{\dagger} \hat{\gamma}^{(\textsf{a})} U_{\bm{k}}\right]_{\overline{\textsf{a}}\textsf{a}}\Big\}\notag \\
\times \mathcal{G}_{\bm{k},\textsf{a}}^R(\epsilon)\mathcal{G}_{\bm{k}+\bm{k}',\overline{\textsf{a}}}^A(\epsilon+\epsilon')\mathcal{G}_{\bm{q}_++\bm{k}',\textsf{a}}^R(\omega_++\epsilon')+c.c.
\end{gather}

In this formula, the curly braces in the second line contain two terms. The first term can be of any sign and corresponds to a diagram with two intersecting arcs in Fig.~\ref{Fig:SII:MainDiagrams}. The second term corresponds to the diagram with a fermion bubble in Fig.~\ref{Fig:SII:MainDiagrams} and is always negative due to the presence of this bubble.

The result for $\Xi^{(\textsf{a})}_{\rm I}(0,0)$, cf. Eq. \eqref{SI:Xi}, after integration over frequencies and simplification has the form
\begin{gather}
	\Xi^{(\textsf{a})}_{\rm I}(0,0)=-\frac{(2\pi)^d}{\int_{\bm{k}}[U_{\bm{k}}^{\dagger}\hat{\gamma}^{(\textsf{a})}U_{\bm{k}}]_{\textsf{aa}}\left|v_{\bm{k}}\right|^{-2}}\int\limits_{\bm{p}_i}\delta(\bm{p}_1{-}\bm{p}_2{+}\bm{p}_3{-}\bm{p}_4)\left|\left(v_{\bm{p}_4} U^{\dag}_{\bm{p}_3}\hat{\gamma}_{\textsf{a}} U_{\bm{p}_2}{-}v_{\bm{p}_2} U^{\dag}_{\bm{p}_3}\hat{\gamma}_{\textsf{a}} U_{\bm{p}_4}\right)_{\bar{\textsf{a}}\textsf{a}}\right|^2\notag \\
\times	\left[U_{\bm{p}_1}^{\dagger} \hat{\gamma}^{(\textsf{a})}U_{\bm{p}_1}\right]_{\textsf{aa}}\frac{\overline{\gamma}_{\textsf{a}}|v_{\bm{p}_1}|^2{+}\overline{\gamma}_{\textsf{a}}|v_{\bm{p}_2}|^2{+}\overline{\gamma}_{\overline{\textsf{a}}}|v_{\bm{p}_3}|^2{+}\overline{\gamma}_{\textsf{a}}|v_{\bm{p}_4}|^2}{(\xi_{\bm{p}_1}{-}\xi_{\bm{p}_2}{-}\xi_{\bm{p}_3}{-}\xi_{\bm{p}_4})^2{+}(\overline{\gamma}_{\textsf{a}}|v_{\bm{p}_1}|^2{+}\overline{\gamma}_{\textsf{a}}|v_{\bm{p}_2}|^2{+}\overline{\gamma}_{\overline{\textsf{a}}}|v_{\bm{p}_3}|^2{+}\overline{\gamma}_{\textsf{a}}|v_{\bm{p}_4}|^2)^2}.
    \label{eq:AppA:Xi1}
\end{gather}

Here we see that the terms coming from the diagrams with arcs and with a bubble in Fig.~\ref{Fig:SII:MainDiagrams} gathered into a complete square, and the final expression turned out to be strictly non-positive, $\Xi^{(\textsf{a})}_{\rm I}(0,0){\leq}0$.

\begin{figure}[h!]
\centerline{\includegraphics[width=1.05\columnwidth]{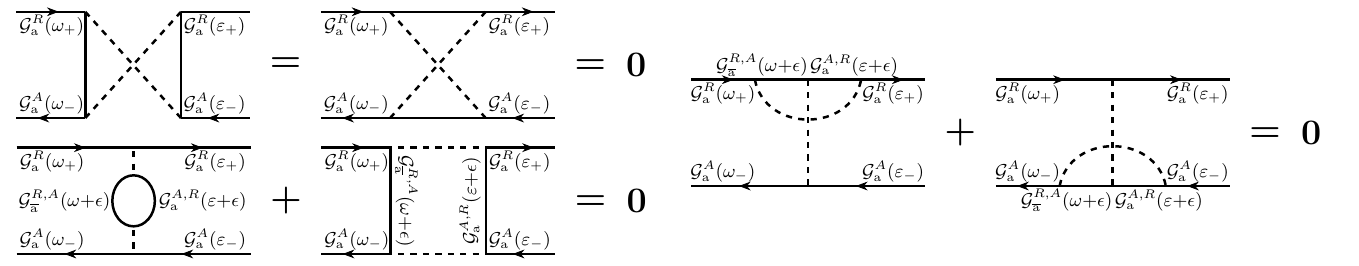}} 
\caption{A way of canceling the diagrams that do not contribute to the self-energy at zero momentum $\bm{Q}$ and zero frequency $\Omega$. The arrangement of matrices $\overline{\mathcal{L}}$ and $\mathcal{L}$ on these diagrams is uniquely determined by $\textsf{a}{=}\textsf{u}$ or $\textsf{a}{=}\textsf{d}$.}
\label{Fig:SII:ZeroDiagrams}
\end{figure}

Now let's look at how all the other diagrams in Fig.~3c in the main text (pale ones) vanish. The contraction mechanism is shown in Fig.~\ref{Fig:SII:ZeroDiagrams}. For example, two diagrams with crosses are equal to zero even with non-zero momentum $\bm{Q}$ and frequency $\Omega$ due to causality: the poles of the Green's functions, which participate in integration over frequencies, cannot be placed on opposite sides of the real axis. The remaining four diagrams are themselves non-zero, but are canceled in pairs after the momentum $\bm{Q}$ and frequency $\Omega$ are equated to zero.

Let us denote these second order diagrammatic insertions in Fig.~\ref{Fig:SII:MainDiagrams} as $\Sigma^{(\textsf{a})}_{\rm{I},1}(\bm{Q},\Omega)$, $\Sigma^{(\textsf{a})}_{\rm{I},2}(\bm{Q},\Omega)$, $\Sigma^{(\textsf{a})}_{\rm{I},3}(\bm{Q},\Omega)$, $\Sigma^{(\textsf{a})}_{\rm{I},4}(\bm{Q},\Omega)$ (bubble, square, crossed arch above, crossed arch below respectively) and write down their values.
\begin{align}
	\Sigma^{(\textsf{a})}_{\rm{I},1}(\bm{Q},\Omega) & =-s_{\textsf{a}}\int_{\bm{q},\omega,\bm{p},\varepsilon}v_{\bm{q}_{+}}v_{\bm{q}_{-}}^*\left[U_{\bm{p}_{+}}^{\dagger} \hat{\gamma}^{(\textsf{a})}U_{\bm{p}_{-}}\right]_{\textsf{aa}}\mathcal{G}_{\bm{q}_{+},\textsf{a}}^R(\omega_{+}) \mathcal{G}_{\bm{q}_{-},\textsf{a}}^A(\omega_{-}) \mathcal{G}_{\bm{p}_{+},\textsf{a}}^R(\varepsilon_{+}) \mathcal{G}_{\bm{p}_{-},\textsf{a}}^A(\varepsilon_{-})\int_{\epsilon, \bm{k}}v_{\bm{p}_+}v_{\bm{p}_-}^*|v_{\bm{p}+\bm{k}}|^2\notag \\
	& \times \left[U_{\bm{q}_{+}}^{\dagger} \hat{\gamma}^{(\textsf{a})} U_{\bm{q}+\bm{k}}\right]_{\textsf{a}\overline{\textsf{a}}}\left[U_{\bm{q}+\bm{k}}^{\dagger} \hat{\gamma}^{(\textsf{a})} U_{\bm{q}_{-}}\right]_{\overline{\textsf{a}}\textsf{a}}\Big\{\mathcal{G}_{\bm{p}+\bm{k},\textsf{a}}^R(\varepsilon+\epsilon)\mathcal{G}_{\bm{q}+\bm{k},\overline{\textsf{a}}}^A(\omega+\epsilon)+\mathcal{G}_{\bm{p}+\bm{k},\textsf{a}}^A(\varepsilon+\epsilon)\mathcal{G}_{\bm{q}+\bm{k},\overline{\textsf{a}}}^R(\omega+\epsilon)\Big\}.
\end{align}
\begin{align}
	\Sigma^{(\textsf{a})}_{\rm{I},2}(\bm{Q},\Omega)& =s_{\textsf{a}}\int_{\bm{q},\omega,\bm{p},\varepsilon}v_{\bm{q}_{+}}v_{\bm{q}_{-}}^*\left[U_{\bm{p}_{+}}^{\dagger} \hat{\gamma}^{(\textsf{a})}U_{\bm{p}_{-}}\right]_{\textsf{aa}}\mathcal{G}_{\bm{q}_{+},\textsf{a}}^R(\omega_{+}) \mathcal{G}_{\bm{q}_{-},\textsf{a}}^A(\omega_{-}) \mathcal{G}_{\bm{p}_{+},\textsf{a}}^R(\varepsilon_{+}) \mathcal{G}_{\bm{p}_{-},\textsf{a}}^A(\varepsilon_{-})\int_{\epsilon, \bm{k}}v_{\bm{p}_+}v_{\bm{p}_-}^*v_{\bm{q}_-}v_{\bm{q}_+}^* \notag \\
	& \times \left[U_{\bm{p}+\bm{k}}^{\dagger} \hat{\gamma}^{(\textsf{a})} U_{\bm{q}+\bm{k}}\right]_{\textsf{a}\overline{\textsf{a}}}\left[U_{\bm{q}+\bm{k}}^{\dagger} \hat{\gamma}^{(\textsf{a})} U_{\bm{p}+\bm{k}}\right]_{\overline{\textsf{a}}\textsf{a}}\Big\{\mathcal{G}_{\bm{p}+\bm{k},\textsf{a}}^R(\varepsilon+\epsilon)\mathcal{G}_{\bm{q}+\bm{k},\overline{\textsf{a}}}^A(\omega+\epsilon)+\mathcal{G}_{\bm{p}+\bm{k},\textsf{a}}^A(\varepsilon+\epsilon)\mathcal{G}_{\bm{q}+\bm{k},\overline{\textsf{a}}}^R(\omega+\epsilon)\Big\}.
\end{align}
\begin{align}
	\Sigma^{(\textsf{a})}_{\rm{I},3}(\bm{Q},\Omega)& =\int_{\bm{q},\omega,\bm{p},\varepsilon}v_{\bm{q}_{+}}v_{\bm{q}_{-}}^*\left[U_{\bm{p}_{+}}^{\dagger} \hat{\gamma}^{(\textsf{a})}U_{\bm{p}_{-}}\right]_{\textsf{aa}}\mathcal{G}_{\bm{q}_{+},\textsf{a}}^R(\omega_{+}) \mathcal{G}_{\bm{q}_{-},\textsf{a}}^A(\omega_{-}) \mathcal{G}_{\bm{p}_{+},\textsf{a}}^R(\varepsilon_{+}) \mathcal{G}_{\bm{p}_{-},\textsf{a}}^A(\varepsilon_{-})\int_{\epsilon, \bm{k}}v_{\bm{p}_+}v_{\bm{p}_-}^*v_{\bm{p}+\bm{k}}v_{\bm{q}_+}^* \notag \\
  & \times  \left[U^{\dagger}_{\bm{p}+\bm{k}}\hat{\gamma}^{(\textsf{a})}U_{\bm{q}+\bm{k}}\right]_{\textsf{a}\overline{\textsf{a}}}\left[U^{\dagger}_{\bm{q}+\bm{k}}\hat{\gamma}^{(\textsf{a})}U_{\bm{q}_-}\right]_{\overline{\textsf{a}}\textsf{a}}\Big\{\mathcal{G}_{\bm{p}+\bm{k},\textsf{a}}^R(\varepsilon+\epsilon)\mathcal{G}_{\bm{q}+\bm{k},\overline{\textsf{a}}}^A(\omega+\epsilon)-\mathcal{G}_{\bm{p}+\bm{k},\textsf{a}}^A(\varepsilon+\epsilon)\mathcal{G}_{\bm{q}+\bm{k},\overline{\textsf{a}}}^R(\omega+\epsilon)\Big\}.
\end{align}
\begin{align}
	\Sigma^{(\textsf{a})}_{\rm{I},4}(\bm{Q},\Omega)& =\int_{\bm{q},\omega,\bm{p},\varepsilon}v_{\bm{q}_{+}}v_{\bm{q}_{-}}^*\left[U_{\bm{p}_{+}}^{\dagger} \hat{\gamma}^{(\textsf{a})}U_{\bm{p}_{-}}\right]_{\textsf{aa}}\mathcal{G}_{\bm{q}_{+},\textsf{a}}^R(\omega_{+}) \mathcal{G}_{\bm{q}_{-},\textsf{a}}^A(\omega_{-}) \mathcal{G}_{\bm{p}_{+},\textsf{a}}^R(\varepsilon_{+}) \mathcal{G}_{\bm{p}_{-},\textsf{a}}^A(\varepsilon_{-})\int_{\epsilon, \bm{k}}v_{\bm{p}_+}v_{\bm{p}_-}^*v_{\bm{q}_-}v_{\bm{p}+\bm{k}}^*\notag \\
   & \times \left[U^{\dagger}_{\bm{q}_+}\hat{\gamma}^{(\textsf{a})}U_{\bm{q}+\bm{k}}\right]_{\textsf{a}\overline{\textsf{a}}}\left[U^{\dagger}_{\bm{q}+\bm{k}}\hat{\gamma}^{(\textsf{a})}U_{\bm{p}+\bm{k}}\right]_{\overline{\textsf{a}}\textsf{a}}\Big\{\mathcal{G}_{\bm{p}+\bm{k},\textsf{a}}^A(\varepsilon+\epsilon)\mathcal{G}_{\bm{q}+\bm{k},\overline{\textsf{a}}}^R(\omega+\epsilon)-\mathcal{G}_{\bm{p}+\bm{k},\textsf{a}}^R(\varepsilon+\epsilon)\mathcal{G}_{\bm{q}+\bm{k},\overline{\textsf{a}}}^A(\omega+\epsilon)\Big\}.
\end{align}
After integration and change of variables, it is easy to verify that
\begin{align}
	\Sigma^{(\textsf{a})}_{\rm{I},2}(0,0)& =-\Sigma^{(\textsf{a})}_{\rm{I},1}(0,0)=\frac{s_{\textsf{a}}}{2\overline{\gamma}_{\textsf{a}}^2}\int_{\bm{q},\bm{p},\bm{k}}|v_{\bm{q}}|^2\left[U_{\bm{p}}^{\dagger} \hat{\gamma}^{(\textsf{a})}U_{\bm{p}}\right]_{\textsf{aa}}\left[U_{\bm{p}+\bm{k}}^{\dagger} \hat{\gamma}^{(\textsf{a})} U_{\bm{q}+\bm{k}}\right]_{\textsf{a}\overline{\textsf{a}}}\left[U_{\bm{q}+\bm{k}}^{\dagger} \hat{\gamma}^{(\textsf{a})} U_{\bm{p}+\bm{k}}\right]_{\overline{\textsf{a}}\textsf{a}}\notag \\
   & \times \frac{\overline{\gamma}_{\overline{\textsf{a}}} |v_{\bm{q}+\bm{k}}|^2{+}\overline{\gamma}_{\textsf{a}}|v_{\bm{p}+\bm{k}}|^2{+}\overline{\gamma}_{\textsf{a}}|v_{\bm{p}}|^2{+}\overline{\gamma}_{\textsf{a}}|v_{\bm{q}}|^2}{(\xi_{\bm{q}+\bm{k}}{+}\xi_{\bm{p}+\bm{k}}{-}\xi_{\bm{p}}{+}\xi_{\bm{q}})^2{+}(\overline{\gamma}_{\overline{\textsf{a}}}|v_{\bm{q}+\bm{k}}|^2{+}\overline{\gamma}_{\textsf{a}}|v_{\bm{p}+\bm{k}}|^2{+}\overline{\gamma}_{\textsf{a}}|v_{\bm{p}}|^2{+}\overline{\gamma}_{\textsf{a}}|v_{\bm{q}}|^2)^2},
   \label{eq:app:A7}
\end{align}
\begin{align}
	\Sigma^{(\textsf{a})}_{\rm{I},3}(\bm{Q},\Omega)& =-\Sigma^{(\textsf{a})}_{\rm{I},4}(\bm{Q},\Omega)=\frac{is_{\textsf{a}}}{2\overline{\gamma}_{\textsf{a}}^2}\int_{\bm{q},\bm{p},\bm{k}}v_{\bm{p}+\bm{k}}v_{\bm{q}}^*\left[U_{\bm{p}}^{\dagger} \hat{\gamma}^{(\textsf{a})}U_{\bm{p}}\right]_{\textsf{aa}}\left[U^{\dagger}_{\bm{p}+\bm{k}}\hat{\gamma}^{(\textsf{a})}U_{\bm{q}+\bm{k}}\right]_{\textsf{a}\overline{\textsf{a}}}\left[U^{\dagger}_{\bm{q}+\bm{k}}\hat{\gamma}^{(\textsf{a})}U_{\bm{q}}\right]_{\overline{\textsf{a}}\textsf{a}}\notag \\
   & \times  \frac{\xi_{\bm{q}+\bm{k}}{+}\xi_{\bm{p}+\bm{k}}{-}\xi_{\bm{p}}{+}\xi_{\bm{q}}}{(\xi_{\bm{q}+\bm{k}}{+}\xi_{\bm{p}+\bm{k}}{-}\xi_{\bm{p}}{+}\xi_{\bm{q}})^2{+}(\overline{\gamma}_{\overline{\textsf{a}}}|v_{\bm{q}+\bm{k}}|^2{+}\overline{\gamma}_{\textsf{a}}|v_{\bm{p}+\bm{k}}|^2{+}\overline{\gamma}_{\textsf{a}}|v_{\bm{p}}|^2{+}\overline{\gamma}_{\textsf{a}}|v_{\bm{q}}|^2)^2}.
\end{align}

\section{Detailed calculation of one of the contributions to the self-energy. \label{App:Detailed}}

To give a clearer idea of the diagrammatic technique used, we present in this section a detailed calculation for a diagram with a bubble connected by dissipative lines with the retarded and advanced branches of the diffuson, see Fig. \ref{Fig:SII:ZeroDiagrams}. First, we rewrite the dissipative action \eqref{eq:SL:ud} as
\begin{equation}
S_{\rm L}=\frac{i (2\pi)^d}{2} \!\!\int\limits_{\bm{p_{j}},t}\! \delta(\bm{p_1}{-}\bm{p_2}{+}\bm{p_3}{-}\bm{p_4})\sum_{\textsf{a},\alpha}\sum_{\nu,\mu{=}0,1}
P_{\mu\nu} \gamma_{\textsf{a},\alpha}\overline{c}_{\bm{p_1}}\tau_\mu[\mathcal{L}^{(\textsf{a},\alpha)}_{\bm{p_2p_1}}]^\dag 
c_{\bm{p_2}}\overline{c}_{\bm{p_3}}\tau_\nu\mathcal{L}^{(\textsf{a},\alpha)}_{\bm{p_3p_4}}
c_{\bm{p_4}}.
\label{eq:App:SL}
\end{equation}
Here we introduce $2{\times}2$ matrix $P^{\mu\nu}{=}\{\{2,1\},\{-1,0\}\}$. $\tau_{0}$ and $\tau_1$ are the identity matrix and the standard $\tau_x$ Pauli matrix, acting in the Keldysh space. Let us also recall the matrix structure of the single-particle Green's functions, $\mathcal{G}_{\bm{p},\textsf{a}}(\varepsilon){=}\mathcal{G}^R_{\bm{p},\textsf{a}}(\varepsilon) \Lambda^{(+)}_\textsf{a} {+} \mathcal{G}^A_{\bm{p},\textsf{a}}(\varepsilon) \Lambda^{(-)}_\textsf{a}$, see Eq.~\eqref{eq:Lambda}. These notations allow us to formulate specific selection rules for diagrams, stated at the beginning of the Appendix~\ref{App:SE}:

(1) the structure of the matrix $P^{\mu\nu}$ prohibits connections of vertices $\Lambda_{\textsf{u}}^{(+)}\tau_{\mu}$ or $\tau_{\mu}\Lambda_{\textsf{d}}^{(-)}$ with $\Lambda_{\textsf{d}}^{(+)}\tau_{\nu}$ or $\tau_{\nu}\Lambda_{\textsf{u}}^{(-)}$ via dissipative lines;

(2) 
Eq.~\eqref{eq:Lambda} leads to $\Lambda_{\textsf{a}}^{(+)}\tau_{\mu}\Lambda_{\textsf{a}}^{(-)}{=}0$;

(3) 
Eq.~\eqref{eq:A:matrices:0} leads to $[\mathcal{L}_{\boldsymbol{p q}}^{(\textsf{a}, \alpha)}]_{\textsf{ud}}{=}[\mathcal{L}_{\boldsymbol{p q}}^{(\textsf{a}, \alpha)}]^{\dag}_{\textsf{du}}{=}0$ and $[\mathcal{L}_{\boldsymbol{p_1p_2}}^{(\textsf{a}, \alpha)}]^{\dag}_{\textsf{uu}}[\mathcal{L}_{\boldsymbol{p_3p_4}}^{(\textsf{a}, \alpha)}]_{\textsf{dd}}{=}[\mathcal{L}_{\boldsymbol{p_1p_2}}^{(\textsf{a}, \alpha)}]^{\dag}_{\textsf{dd}}[\mathcal{L}_{\boldsymbol{p_3p_4}}^{(\textsf{a}, \alpha)}]_{\textsf{uu}}{=}0$;

(4) the poles of the Green's functions in the integrals over frequencies must be located on different sides of the integration contour, otherwise these integrals will give zero.

\begin{figure}[h!]
\centerline{\includegraphics[width=0.6\columnwidth]{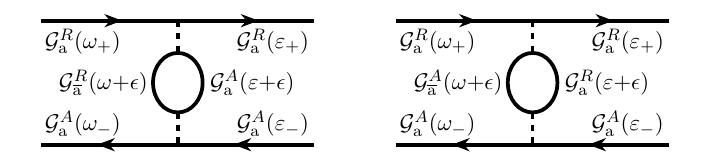}} 
\caption{Two possible arrangements of indices on the bubble diagram $\Sigma^{(\textsf{a})}_{\rm bub}(\bm{Q},\Omega)$. The arrangement of matrices $\mathcal{L}^\dag$ and $\mathcal{L}$ on these diagrams is uniquely determined by $\textsf{a}{=}\textsf{u}$ or $\textsf{a}{=}\textsf{d}$.}
\label{Fig:SII:Bubble}
\end{figure}

According to these rules, there are two ways to arrange the indices on the bubble diagram, see Fig.~\ref{Fig:SII:Bubble}.

To be specific, let's set $\textsf{a}{=}\textsf{u}$. The sum of inserted diagrams on Fig.~\ref{Fig:SII:Bubble} equals
\begin{gather}
\xi^{(\textsf{u})}_{\rm{I},1}(\boldsymbol{Q},\Omega,\boldsymbol{q},\omega,\boldsymbol{p},\varepsilon)\big[\Lambda_u^{(+)}\big]^{\beta\beta'}\big[\Lambda_u^{(-)}\big]^{\gamma\gamma'}=-\frac{1}{4}\int_{\epsilon, \boldsymbol{k}}\sum_{\textsf{a},\alpha}\sum_{\nu,\mu}\sum_{\textsf{a}',\alpha'}\sum_{\nu',\mu'}P^{\mu \nu}\gamma_{\textsf{a},\alpha}P^{\mu'\nu'}\gamma_{\textsf{a}',\alpha'}
 \big[\Lambda_u^{(+)} \tau_\nu  \Lambda_u^{(+)}\big]^{\beta\beta'}\big[\Lambda_u^{(-)} \tau_{\mu'} \Lambda_u^{(-)}\big]^{\gamma\gamma'}
 \notag \\
 \times
 [\mathcal{L}_{(\boldsymbol{q}+\boldsymbol{k})(\boldsymbol{p}+\boldsymbol{k})}^{(\textsf{a},\alpha)}]^{\dag}_{\textsf{ud}}[\mathcal{L}_{\boldsymbol{q}_+\boldsymbol{p}_+}^{(\textsf{a},\alpha)}]_{\textsf{uu}}[\mathcal{L}_{\boldsymbol{q}_-\boldsymbol{p}_-}^{(\textsf{a}',\alpha')}]^{\dag}_{\textsf{uu}}[\mathcal{L}_{(\boldsymbol{q}+\boldsymbol{k})(\boldsymbol{p}+\boldsymbol{k})}^{(\textsf{a}',\alpha')}]_{\textsf{du}}
\sum_{\delta}\Big\{\mathcal{G}_{\boldsymbol{q}+\boldsymbol{k},\textsf{d}}^R(\omega{+}\epsilon)\mathcal{G}_{\boldsymbol{p}+\boldsymbol{k},\textsf{u}}^A(\varepsilon{+}\epsilon)\big[\Lambda_u^{(-)} \tau_\mu \Lambda_d^{(+)}\tau_{\nu'}\big]^{\delta\delta}
\notag \\
+\mathcal{G}_{\boldsymbol{q}+\boldsymbol{k},\textsf{d}}^A(\omega{+}\epsilon)\mathcal{G}_{\boldsymbol{p}+\boldsymbol{k},\textsf{u}}^R(\varepsilon{+}\epsilon)\big[\Lambda_u^{(+)} \tau_\mu \Lambda_d^{(-)}\tau_{\nu'}\big]^{\delta\delta}\Big\} =
		-\big[\Lambda_u^{(+)}\big]^{\beta\beta'}\big[\Lambda_u^{(-)}\big]^{\gamma\gamma'}\int_{\epsilon, \boldsymbol{k}}v_{\boldsymbol{p}_+}v_{\boldsymbol{p}_-}^*|v_{\boldsymbol{p}+\boldsymbol{k}}|^2\left[U_{\boldsymbol{q}_{+}}^{\dagger} \hat{\gamma}^{(\textsf{u})} U_{\boldsymbol{q}+\boldsymbol{k}}\right]_{\textsf{ud}}\notag \\
        \times \left[U_{\boldsymbol{q}+\boldsymbol{k}}^{\dagger} \hat{\gamma}^{(\textsf{u})} U_{\boldsymbol{q}_{-}}\right]_{\textsf{du}}
		\Big\{\mathcal{G}_{\boldsymbol{p}+\boldsymbol{k},\textsf{u}}^R(\varepsilon{+}\epsilon)\mathcal{G}_{\boldsymbol{q}+\boldsymbol{k},\textsf{d}}^A(\omega{+}\epsilon)+\mathcal{G}_{\boldsymbol{p}+\boldsymbol{k},\textsf{u}}^A(\varepsilon{+}\epsilon)\mathcal{G}_{\boldsymbol{q}+\boldsymbol{k},\textsf{d}}^R(\omega{+}\epsilon)\Big\}.
	\end{gather}

Here Dirac delta function with the law of conservation of momentum is removed by integration over the fourth momentum. Using the definition \eqref{SI:Sigma}, we find
\begin{equation}
	\Sigma^{(\textsf{u})}_{\rm{I},1}(\boldsymbol{Q},\Omega)=\int_{\boldsymbol{q},\omega,\boldsymbol{p},\varepsilon}v_{\boldsymbol{q}_{+}}v_{\boldsymbol{q}_{-}}^*\left[U_{\boldsymbol{p}_{+}}^{\dagger} \hat{\gamma}^{(\textsf{u})}U_{\boldsymbol{p}_{-}}\right]_{\textsf{uu}}\mathcal{G}_{\boldsymbol{q}_{+},\textsf{u}}^R(\omega_{+}) \mathcal{G}_{\boldsymbol{q}_{-},\textsf{u}}^A(\omega_{-}) \mathcal{G}_{\boldsymbol{p}_{+},\textsf{u}}^R(\varepsilon_{+}) \mathcal{G}_{\boldsymbol{p}_{-},\textsf{u}}^A(\varepsilon_{-})\xi^{(\textsf{u})}_{\rm{I},1}(\boldsymbol{Q},\Omega,\boldsymbol{q},\omega,\boldsymbol{p},\varepsilon).
\end{equation}

After the integration over frequencies ($\epsilon$, $\omega$, $\varepsilon$) this expression takes form
\begin{align}
    \Sigma^{(\textsf{u})}_{\rm{I},1}(\boldsymbol{Q},\Omega) & =i\int_{\boldsymbol{q},\boldsymbol{p},\boldsymbol{k}}\frac{v_{\boldsymbol{q}_{+}}v_{\boldsymbol{q}_{-}}^*v_{\boldsymbol{p}_+}v_{\boldsymbol{p}_-}^*|v_{\boldsymbol{p}+\boldsymbol{k}}|^2\left[U_{\boldsymbol{p}_{+}}^{\dagger} \hat{\gamma}^{(\textsf{u})}U_{\boldsymbol{p}_{-}}\right]_{\textsf{uu}}\left[U_{\boldsymbol{q}_{+}}^{\dagger} \hat{\gamma}^{(\textsf{u})} U_{\boldsymbol{q}+\boldsymbol{k}}\right]_{\textsf{ud}} \left[U_{\boldsymbol{q}+\boldsymbol{k}}^{\dagger} \hat{\gamma}^{(\textsf{u})}U_{\boldsymbol{q}_{-}}\right]_{\textsf{du}}}{[\Omega{-}\xi_{\boldsymbol{q}_+}{+}\xi_{\boldsymbol{q}_-}{+}i\overline{\gamma}_{\textsf{u}}(|v_{\boldsymbol{q}_+}|^2{+}|v_{\boldsymbol{q}_-}|^2)][\Omega{-}\xi_{\boldsymbol{p}_+}{+}\xi_{\boldsymbol{p}_-}{+}i\overline{\gamma}_{\textsf{u}}(|v_{\boldsymbol{p}_+}|^2{+}|v_{\boldsymbol{p}_-}|^2)]} \notag \\
 & \times    \bigg\{\frac{1}{\Omega{+}\xi_{\boldsymbol{q}+\boldsymbol{k}}{+}\xi_{\boldsymbol{p}+\boldsymbol{k}}{-}\xi_{\boldsymbol{p}_+}{+}\xi_{\boldsymbol{q}_-}{+}i(\overline{\gamma}_{\textsf{d}}|v_{\boldsymbol{q}+\boldsymbol{k}}|^2{+}\overline{\gamma}_{\textsf{u}}|v_{\boldsymbol{p}+\boldsymbol{k}}|^2{+}\overline{\gamma}_{\textsf{u}}|v_{\boldsymbol{p}_+}|^2{+}\overline{\gamma}_{\textsf{u}}|v_{\boldsymbol{q}_-}|^2)}\\
		& -\frac{1}{-\Omega{+}\xi_{\boldsymbol{q}+\boldsymbol{k}}{+}\xi_{\boldsymbol{p}+\boldsymbol{k}}{-}\xi_{\boldsymbol{p}_-}{+}\xi_{\boldsymbol{q}_+}{-}i(\overline{\gamma}_{\textsf{d}} |v_{\boldsymbol{q}+\boldsymbol{k}}|^2{+}\overline{\gamma}_{\textsf{u}}|v_{\boldsymbol{p}+\boldsymbol{k}}|^2{+}\overline{\gamma}_{\textsf{u}}|v_{\boldsymbol{p}_-}|^2{+}\overline{\gamma}_{\textsf{u}}|v_{\boldsymbol{q}_+}|^2)}\bigg\}.
	\end{align}
For zero momentum $\boldsymbol{Q}{=}0$ and zero frequency $\Omega{=}0$, we have
\begin{align}
    \Sigma^{(\textsf{u})}_{\rm{I},1}(0,0) & =-\frac{1}{2\overline{\gamma}_{\textsf{u}}^2}
\int_{\boldsymbol{q},\boldsymbol{p},\boldsymbol{k}}|v_{\boldsymbol{p}+\boldsymbol{k}}|^2\left[U_{\boldsymbol{p}}^{\dagger} \hat{\gamma}^{(\textsf{u})}U_{\boldsymbol{p}}\right]_{\textsf{uu}}\left[U_{\boldsymbol{q}}^{\dagger}\hat{\gamma}^{(\textsf{u})} U_{\boldsymbol{q}+\boldsymbol{k}}\right]_{\textsf{ud}}\left[U_{\boldsymbol{q}+\boldsymbol{k}}^{\dagger} \hat{\gamma}^{(\textsf{u})}U_{\boldsymbol{q}}\right]_{\textsf{du}} \notag \\
		& \times \frac{\overline{\gamma}_{\textsf{d}} |v_{\boldsymbol{q}+\boldsymbol{k}}|^2{+}\overline{\gamma}_{\textsf{u}}|v_{\boldsymbol{p}+\boldsymbol{k}}|^2{+}\overline{\gamma}_{\textsf{u}}|v_{\boldsymbol{p}}|^2{+}\overline{\gamma}_{\textsf{u}}|v_{\boldsymbol{q}}|^2}{(\xi_{\boldsymbol{q}+\boldsymbol{k}}{+}\xi_{\boldsymbol{p}+\boldsymbol{k}}{-}\xi_{\boldsymbol{p}}{+}\xi_{\boldsymbol{q}})^2{+}(\overline{\gamma}_{\textsf{d}}|v_{\boldsymbol{q}+\boldsymbol{k}}|^2{+}\overline{\gamma}_{\textsf{u}}|v_{\boldsymbol{p}+\boldsymbol{k}}|^2{+}\overline{\gamma}_{\textsf{u}}|v_{\boldsymbol{p}}|^2{+}\overline{\gamma}_{\textsf{u}}|v_{\boldsymbol{q}}|^2)^2} ,
	\label{Sigma3}
\end{align}
which coincides with Eq. \eqref{eq:app:A7}.

\section{Insertion of an opposite band diffuson inside the ladder. \label{App:InsertionDiffusons}}

\begin{figure}[h!]
\centerline{\includegraphics[width=0.6\columnwidth]{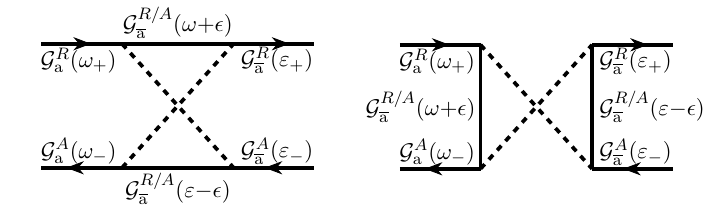}} 
\caption{Arrangement of indices on diagrams that make a real contribution to $\Upsilon_{\textsf{a}\bar{\textsf{a}}}(0,0)$. The arrangement of matrices $\overline{\mathcal{L}}$ and $\mathcal{L}$ on these diagrams is uniquely determined by $\textsf{a}{=}\textsf{u}$ or $\textsf{a}{=}\textsf{d}$.}
\label{Fig:SIII:MainDiagrams}
\end{figure}

Now let's move on to the self-energy contribution $\Xi^{(\textsf{a})}_{\rm II}$, which is obtained as a result of combining diffusons from the different bands into a single ladder (see Fig. \ref{Figure:SE:diagrams2}). Firstly, we need to calculate transition diagrams that are needed to connect two different diffusons. Let us denote them as $\Upsilon_{\textsf{a}\bar{\textsf{a}}}(\bm{Q},\Omega)$ and consider all possible variants of diagrams with a minimum number of diffusion lines (namely, two), which have $\textsf{u}$-ends on one side and $\textsf{d}$-ends on the other (see Fig.~\ref{Fig:SIII:MainDiagrams}).

Remembering the definition \eqref{SI:Sigma} (but now we need to take into account that there are different diffusons on both sides), we get
\begin{equation}
\begin{gathered}
	\Upsilon_{\textsf{a}\bar{\textsf{a}}}(\bm{Q},\Omega)=-\frac{4\bar{\gamma}_{\textsf{a}}^2}{\int_{\bm{k}}[U_{\bm{k}}^{\dagger}\hat{\gamma}^{(\textsf{a})}U_{\bm{k}}]_{\textsf{aa}}\left|v_{\bm{k}}\right|^{-2}}\int_{\bm{q},\omega,\bm{p},\varepsilon}v_{\bm{q}_{+}}v_{\bm{q}_{-}}^*\left[U_{\bm{p}_{+}}^{\dagger} \hat{\gamma}^{(\bar{\textsf{a}})}U_{\bm{p}_{-}}\right]_{\bar{\textsf{aa}}}\mathcal{G}_{\bm{q}_{+},\textsf{a}}^R(\omega_{+}) \mathcal{G}_{\bm{q}_{-},\textsf{a}}^A(\omega_{-}) \mathcal{G}_{\bm{p}_{+},\bar{\textsf{a}}}^R(\varepsilon_{+})\mathcal{G}_{\bm{p}_{-},\bar{\textsf{a}}}^R(\varepsilon_{-})\\
    \int_{\epsilon,\bm{k}} v_{\bm{q}+\bm{k}}v_{\bm{p}_+}v_{\bm{p}_-}^*\left[U_{\bm{q}_+}^{\dagger} \hat{\gamma}^{(\overline{\textsf{a}})} U_{\bm{p}-\bm{k}}\right]_{\textsf{a}\overline{\textsf{a}}}\left(v_{\bm{p}-\bm{k}}^*\left[U_{\bm{q}+\bm{k}}^{\dagger} \hat{\gamma}^{(\overline{\textsf{a}})} U_{\bm{q}_-}\right]_{\overline{\textsf{a}}\textsf{a}}-v_{\bm{q}+\bm{k}}^*\left[U_{\bm{p}-\bm{k}}^{\dagger} \hat{\gamma}^{(\overline{\textsf{a}})} U_{\bm{q}_-}\right]_{\overline{\textsf{a}}\textsf{a}}\right)\\
	\left\{\mathcal{G}_{\bm{q}+\bm{k},\overline{\textsf{a}}}^R(\omega+\epsilon)\mathcal{G}_{\bm{p}-\bm{k},\overline{\textsf{a}}}^R(\varepsilon-\epsilon)+\mathcal{G}_{\bm{q}+\bm{k},\overline{\textsf{a}}}^A(\omega+\epsilon)\mathcal{G}_{\bm{p}-\bm{k},\overline{\textsf{a}}}^A(\varepsilon-\epsilon)\right\}.
\end{gathered}
\end{equation}

After integration over frequencies, we obtain
\begin{equation}
\begin{gathered}
	\Upsilon_{\bar{\textsf{a}}\textsf{a}}(0,0)=-\frac{\bar{\gamma}_{\textsf{a}}(2\pi)^d}{\overline{\gamma}_{\overline{\textsf{a}}}\int_{\bm{k}}[U_{\bm{k}}^{\dagger}\hat{\gamma}^{(\textsf{a})}U_{\bm{k}}]_{\textsf{aa}}\left|v_{\bm{k}}\right|^{-2}}\int_{\bm{p}_i}\delta(\bm{p}_1{-}\bm{p}_2{+}\bm{p}_3{-}\bm{p}_4)\left|\left(v_{\bm{p}_4} U^{\dag}_{\bm{p}_3}\hat{\gamma}_{\textsf{a}} U_{\bm{p}_2}{-}v_{\bm{p}_2} U^{\dag}_{\bm{p}_3}\hat{\gamma}_{\textsf{a}} U_{\bm{p}_4}\right)_{\bar{\textsf{a}}\textsf{a}}\right|^2\\
	\left[U_{\bm{p}_1}^{\dagger}\hat{\gamma}^{(\textsf{a})}U_{\bm{p}_1}\right]_{\textsf{aa}}\frac{\overline{\gamma}_{\textsf{a}}|v_{\bm{p}_1}|^2{+}\overline{\gamma}_{\textsf{a}}|v_{\bm{p}_2}|^2{+}\overline{\gamma}_{\overline{\textsf{a}}}|v_{\bm{p}_3}|^2{+}\overline{\gamma}_{\textsf{a}}|v_{\bm{p}_4}|^2}{(\xi_{\bm{p}_1}{-}\xi_{\bm{p}_2}{-}\xi_{\bm{p}_3}{-}\xi_{\bm{p}_4})^2{+}(\overline{\gamma}_{\textsf{a}}|v_{\bm{p}_1}|^2{+}\overline{\gamma}_{\textsf{a}}|v_{\bm{p}_2}|^2{+}\overline{\gamma}_{\overline{\textsf{a}}}|v_{\bm{p}_3}|^2{+}\overline{\gamma}_{\textsf{a}}|v_{\bm{p}_4}|^2)^2}.
	\end{gathered}
\end{equation}

We find, unexpectedly, that 
$\Upsilon_{\bar{\textsf{a}}\textsf{a}}(0,0){=}(\bar{\gamma}_{\textsf{a}}/\bar{\gamma}_{\bar{\textsf{a}}}) \Xi^{(\textsf{a})}_{\rm I}(0,0)$. The resulting $\textsf{a}-$diffuson self-energy contribution is $\Xi^{(\textsf{a})}_{\rm II}(0,0){=}[\tau_{\textsf{a}}\tau_{\bar{\textsf{a}}}(D^{(\bar{\textsf{a}})}_{jl}Q_jQ_l{-}i\Omega)]^{-1}$.

\begin{figure}[h!]
\centerline{\includegraphics[width=1\columnwidth]{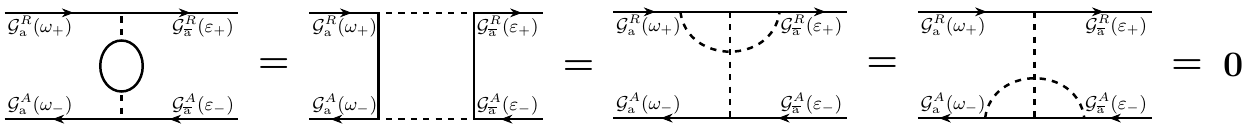}} 
\caption{Connecting diagrams for diffusons of different bands, which turned out to be equal to zero due to causality. The arrangement of matrices $\overline{\mathcal{L}}$ and $\mathcal{L}$ on these diagrams is uniquely determined by $\textsf{a}{=}\textsf{u}$ or $\textsf{a}{=}\textsf{d}$.}
\label{Fig:SIII:ZeroDiagrams}
\end{figure}

Now let's look at how all the other diagrams in Fig.~3e in the main text (pale ones) vanish. They are shown in detail in Fig.~\ref{Fig:SIII:ZeroDiagrams}. These diagrams are equal to zero even with non-zero momentum $\bm{Q}$ and frequency $\Omega$ due to causality: the poles of the Green's functions, which participate in integration over frequencies, cannot be placed on opposite sides of the real axis.

\section{Recombination. \label{App:Recombination}}

We now analyze the last contribution to the self-energy, $\Xi^{(\textsf{a})}_{\rm III}$. It is related with calculation of the Hartree-Fock diagrams (see Fig.~\ref{Fig:SCBA}) taking into account fermion density deviations $\delta n_{\textsf{a}}$ from the dark state. Until now, we have used exclusively the dark state Green's functions and considered the contributions to the self-energy of second order in $\overline{\gamma}_{\textsf{a}}$. Now we are interested in the contribution of the first order in $\overline{\gamma}_{\textsf{a}}$, which is also proportional to the $\delta n_{\textsf{a}}$, since we are going to take into account the deviation of the Green's functions from the dark state. The calculation is carried out in terms of the Keldysh component of the Green's function, so we introduce the notation
\begin{equation}
	\delta \mathcal{G}_{\bm{p}}^K=i\begin{pmatrix}
		\delta u_{\bm{p},} & \delta \eta_{\bm{p}} \\
		\delta \eta_{\bm{p}}^* & \delta d_{\bm{p}}
	\end{pmatrix}.\label{eq:SIV:deltaG}
\end{equation}

Here we use the deviations of the particle distribution functions in the upper and lower bands, $\delta u_{\bm{p}}$ and $\delta d_{\bm{p}}$. They are related to the density deviations in a simple way,
\begin{equation}
	\delta n_{\textsf{u}}=\int_{\bm{p}}\delta u_{\bm{p}},\quad\delta n_{\textsf{d}}=\int_{\bm{p}}\delta d_{\bm{p}}.
\end{equation}

In this notation, after calculating the Hartree-Fock diagrams with shifted Green's functions \eqref{eq:SIV:deltaG}, for the upper band we get the result:
\begin{equation}
\begin{gathered}
\Sigma^{(\textsf{u})}_{\rm III}(0,0)=i\int_{\bm{q},\bm{p},\omega}|v_{\bm{q}}|^2\left[U_{\bm{q}}^{\dagger} \hat{\gamma}^{(\textsf{u})}U_{\bm{q}}\right]_{\textsf{uu}}\mathcal{G}_{\bm{q},\textsf{u}}^R(\omega) \mathcal{G}_{\bm{q},\textsf{u}}^A(\omega)\bigg\{\left[\mathcal{G}_{\bm{q},\textsf{u}}^R(\omega)+\mathcal{G}_{\bm{q},\textsf{u}}^A(\omega)\right]\\
\bigg(\left[v_{\bm{q}}v_{\bm{p}}^*[U_{\bm{q}}^{\dagger}\hat{\gamma}^{(\textsf{u})}U_{\bm{p}}]_{\textsf{uu}}-v_{\bm{p}}v_{\bm{q}}^*[U_{\bm{p}}^{\dagger}\hat{\gamma}^{(\textsf{u})}U_{\bm{q}}]_{\textsf{uu}}\right]\delta u_{\bm{p}}+\left[v_{\bm{q}}v_{\bm{p}}^*[U_{\bm{q}}^{\dagger}\hat{\gamma}^{(\textsf{u})}U_{\bm{p}}]_{\textsf{ud}}\delta \eta_{\bm{p}}^*-v_{\bm{p}}v_{\bm{q}}^*[U_{\bm{p}}^{\dagger}\hat{\gamma}^{(\textsf{u})}U_{\bm{q}}]_{\textsf{du}}\delta \eta_{\bm{p}}\right]\bigg)+\\
+\left[\mathcal{G}_{\bm{q},\textsf{u}}^R(\omega)-\mathcal{G}_{\bm{q},\textsf{u}}^A(\omega)\right]\bigg(\left[|v_{\bm{q}}|^2[U_{\bm{p}}^{\dagger}\hat{\gamma}^{(\textsf{u})}U_{\bm{p}}]_{\textsf{uu}}-|v_{\bm{p}}|^2[U_{\bm{q}}^{\dagger}\hat{\gamma}^{(\textsf{u})}U_{\bm{q}}]_{\textsf{uu}}\right]\delta u_{\bm{p}}+\\
+\left[ |v_{\bm{q}}|^2[U_{\bm{p}}^{\dagger}\hat{\gamma}^{(\textsf{u})}U_{\bm{p}}]_{\textsf{du}}\delta\eta_{\bm{p}}+ |v_{\bm{q}}|^2[U_{\bm{p}}^{\dagger}\hat{\gamma}^{(\textsf{u})}U_{\bm{p}}]_{\textsf{ud}}\delta\eta_{\bm{p}}^*\right]+\left[|v_{\bm{p}}|^2[U_{\bm{q}}^{\dagger}\hat{\gamma}^{(\textsf{d})}U_{\bm{q}}]_{\textsf{uu}}+|v_{\bm{q}}|^2[U_{\bm{p}}^{\dagger}\hat{\gamma}^{(\textsf{u})}U_{\bm{p}}]_{\textsf{dd}}\right]\delta d_{\bm{p}}\bigg)\bigg\}.
\end{gathered}\label{eq:SIV:Sigma}
\end{equation}

This expression can be simplified using the property of the distribution functions obtained in previous work \cite{Nosov2023}. In that article, the formulas were obtained for a specific model, and here we present their generalization:
\begin{equation}
	\begin{gathered}
		\delta u_{\bm{p}}= \frac{\left[U_{\bm{p}}^{\dagger}\hat{\gamma}^{(\textsf{u})}U_{\bm{p}}\right]_{\textsf{uu}}}{\overline{\gamma}_{\textsf{u}}|v_{\bm{p}}|^2}\int_{\bm{k}}|v_{\bm{k}}|^2\delta u_{\bm{k}},\quad\delta d_{\bm{p}}= \frac{\left[U_{\bm{p}}^{\dagger}\hat{\gamma}^{(\textsf{d})}U_{\bm{p}}\right]_{\textsf{dd}}}{\overline{\gamma}_{\textsf{d}}|v_{\bm{p}}|^2}\int_{\bm{k}}|v_{\bm{k}}|^2\delta d_{\bm{k}}.
	\end{gathered} \label{eq:SIV:distribution}
\end{equation}

Similar relations can be written for off-diagonal elements, $\delta \eta_{\bm{p}}$. Using them, it can be shown that the terms of the Eq. \eqref{eq:SIV:Sigma} with $\delta \eta_{\bm{p}}$ and $\delta \eta_{\bm{p}}^*$ have a higher order in $\bar{\gamma}_\textsf{a}$ than the terms proportional to the diagonal elements, $\delta u_{\bm{p}}$ and $\delta d_{\bm{p}}$. Therefore, they will not be included in the final answer.

Simplifying the expression \eqref{eq:SIV:Sigma} using formulas \eqref{eq:SIV:distribution}, we get
\begin{equation}
\begin{gathered}
	\Xi^{(\textsf{u})}_{\rm III}(0,0)=-\frac{2\int_{\bm{p}}\left|v_{\bm{p}}\right|^{-2}\left(\overline{\gamma}_{\textsf{d}}[U_{\bm{p}}^{\dagger} \hat{\gamma}^{(\textsf{u})}U_{\bm{p}}]_{\textsf{uu}}[U_{\bm{p}}^{\dagger} \hat{\gamma}^{(\textsf{d})}U_{\bm{p}}]_{\textsf{uu}}+\overline{\gamma}_{\textsf{u}}[U_{\bm{p}}^{\dagger}\hat{\gamma}^{(\textsf{u})}U_{\bm{p}}]_{\textsf{dd}}[U_{\bm{p}}^{\dagger}\hat{\gamma}^{(\textsf{d})}U_{\bm{p}}]_{\textsf{dd}}\right)}{\int_{\bm{k}}[U_{\bm{k}}^{\dagger}\hat{\gamma}^{(\textsf{u})}U_{\bm{k}}]_{\textsf{uu}}\left|v_{\bm{k}}\right|^{-2}\int_{\bm{q}}[U_{\bm{q}}^{\dagger}\hat{\gamma}^{(\textsf{d})}U_{\bm{q}}]_{\textsf{dd}}\left|v_{\bm{q}}\right|^{-2}}\delta n_{\textsf{d}}.
\end{gathered}
\end{equation}

For the lower band everything is similar.

Therefore, to the lowest order of density deviations and $\bar{\gamma}_\textsf{a}$, we find $\Xi^{(\textsf{a})}_{\rm III}(0,0){=}{-}\beta s_{\textsf{a}}\delta n_{\bar{\textsf{a}}}$ with
$\beta{=}\beta_{\textsf{u}}{+}\beta_{\textsf{d}}$, where
\begin{equation}
\beta_{\textsf{a}}=\frac{2\overline{\gamma}_{\textsf{a}} \int_{\bm{p}}[U_{\bm{p}}^{\dagger} \hat{\gamma}_{\textsf{u}}U_{\bm{p}}]_{\textsf{aa}}[U_{\bm{p}}^{\dagger} \hat{\gamma}_{\textsf{d}}U_{\bm{p}}]_{\textsf{aa}}|v_{\bm{p}}|^{-2}}{\int_{\bm{k}}[U_{\bm{k}}^{\dagger}\hat{\gamma}_{\textsf{u}}U_{\bm{k}}]_{\textsf{uu}}|v_{\bm{k}}|^{-2}\int_{\bm{q}}[U_{\bm{q}}^{\dagger}\hat{\gamma}_{\textsf{d}}U_{\bm{q}}]_{\textsf{dd}}|v_{\bm{q}}|^{-2}}.
\end{equation}

\section{Estimation of $|1/\tau_{\textsf{u}}|$ for the specific model of a topological insulator. \label{App:Estimation}}

Let's calculate the leading contribution to the self-energy $|1/\tau_{\textsf{a}}|$ for the model of a two-dimensional topological insulator from Ref.~\cite{Tonielli2020}, in which $\xi_{\bm{p}}{=}p^2{+}m^2$, $v_{\bm{p}}{=}\sqrt{\xi_{\bm{p}}}$, $U_{\bm{p}}{=}(p_x{-}i p_y\sigma_z {-}i m \sigma_y)/\sqrt{\xi_{\bm{p}}}$, and $\bar{\gamma}_{\textsf{a}}{\equiv}\bar{\gamma}$. The definition Eq.~\eqref{eq:AppA:Xi1} after the substitution $\bm{p}_1{=}\bm{p}$, $\bm{p}_2{=}\bm{k}$, $\bm{p}_3{=}\bm{k}{+}\bm{Q}$ and integration over $\bm{p}_4$ takes the form
\begin{equation}
	\begin{gathered}
		\frac{1}{|\tau_{\textsf{a}}|}=\frac{2\bar{\gamma}^3m^2}{n^2\int_{\bm{k}}\left|v_{\bm{k}}\right|^{-2}}\int_{\bm{p},\bm{k},\bm{Q}}\left(\frac{\xi_{\bm{p}+\bm{Q}}\bm{Q}^2}{\xi_{\bm{k}}\xi_{\bm{k}+\bm{Q}}}{-}\frac{\bm{Q}(\bm{k}{-}\bm{p})}{\xi_{\bm{k}+\bm{Q}}}\right)
		\frac{\xi_{\bm{p}}{+}\xi_{\bm{k}}{+}\xi_{\bm{p}{+}\bm{Q}}{+}\xi_{\bm{k}+\bm{Q}}}{(\xi_{\bm{p}}{-}\xi_{\bm{k}}{-}\xi_{\bm{p}+\bm{Q}}{-}\xi_{\bm{k}+\bm{Q}})^2{+}\bar{\gamma}^2(\xi_{\bm{p}}{+}\xi_{\bm{k}}{+}\xi_{\bm{p}+\bm{Q}}{+}\xi_{\bm{k}+\bm{Q}})^2}.
	\end{gathered}
\end{equation}
Here the number of particles $n{=}\int_{\bm{k}}1$ is introduced. Note that in this model $\bar{\gamma}_{\textsf{a}}{=} \int_p [U^\dag_{\bm{p}} \hat\gamma_{\textsf{a}} U_{\bm{p}}]_{\textsf{a}\textsf{a}}{\equiv}\gamma n{\equiv}\bar{\gamma}$. Now we substitute $m\bm{x}{=}\bm{k}$, $m\bm{y}{=}\bm{p}{+}\bm{Q}$, $m\bm{z}{=}{-}\bm{k}{-}\bm{Q}$ and rewrite the integral as
\begin{equation}
	\begin{gathered}
		\frac{1}{|\tau_{\textsf{a}}|}=\frac{\bar{\gamma}^3m^6n^{-2}}{\int_{\bm{k}}\left|v_{\bm{k}}\right|^{-2}}\int_{\bm{x},\bm{y},\bm{z}}\left(\frac{2(1{+}y^2)(x^2{+}\bm{x}\bm{z})}{(1{+}x^2)(1{+}z^2)}{-}\frac{z^2{+}\bm{x}\bm{y}{+}\bm{x}\bm{z}{+}\bm{y}\bm{z}}{(1{+}z^2)}\right)\frac{2{+}x^2{+}y^2{+}z^2{+}\bm{x}\bm{y}{+}\bm{x}\bm{z}{+}\bm{y}\bm{z}}{(1{-}\bm{x}\bm{y}{-}\bm{x}\bm{z}{-}\bm{y}\bm{z})^2{+}\bar{\gamma}^2(2{+}x^2{+}y^2{+}z^2{+}\bm{x}\bm{y}{+}\bm{x}\bm{z}{+}\bm{y}\bm{z})^2}.
	\end{gathered}\label{eq:AppD:tau}
\end{equation}
The main part of the integral is collected at $|z|{\ll}|x|,|y|$. Let us expand the integrand, move to polar coordinates, and integrate over the angle.
\begin{equation}
	\begin{gathered}
		\frac{1}{|\tau_{\textsf{a}}|}\approx\frac{\bar{\gamma}^3m^6n^{-2}}{\int_{\bm{k}}\left|v_{\bm{k}}\right|^{-2}}\int_{\bm{x},\bm{y},\bm{z}}\frac{y^2}{1{+}z^2}\frac{x^2+y^2}{(\bm{x}\bm{y})^2{+}\bar{\gamma}^2(x^2{+}y^2)^2}=\frac{\bar{\gamma}^3m^6n^{-2}}{(2\pi)^2}\iint\frac{xy\,dx\,dy}{\bar{\gamma}\sqrt{\frac{x^2y^2}{(x^2{+}y^2)^2}{+}\bar{\gamma}^2}}.
	\end{gathered}
\end{equation}
Here, in accordance with the definition of the number of particles $n$, the integral over the momenta is carried out along a circle of radius $\sqrt{4\pi n}$ (or $\sqrt{4\pi n}/m$ for dimensionless variables $x$ and $y$). Now, after expanding in $\bar{\gamma}$, we can take the integral exactly and get the final answer.
\begin{equation}
	\begin{gathered}
		\frac{1}{|\tau_{\textsf{a}}|}\approx\frac{\bar{\gamma}^2m^6}{(2\pi)^2n^2}\iint\limits_0^{\sqrt{4\pi n}/m} dx\,dy\,(x^2{+}y^2)=\frac{8\bar{\gamma}^2m^2}{3}.
	\end{gathered}
\end{equation}

Now we consider a one-dimensional analog of this model with the same $\xi_{p}{=}p^2{+}m^2$, $v_{p}{=}\sqrt{\xi_{p}}$, and $\bar{\gamma}_{\textsf{a}}{\equiv}\bar{\gamma}$, but $U_{p}{=}(p{-}i m \sigma_y)/\sqrt{\xi_{p}}$. Eq.~\eqref{eq:AppD:tau} for $d{=}1$ after some symmetry-based simplifications takes the form
\begin{equation}
	\begin{gathered}
		\frac{1}{|\tau_{\textsf{a}}|}=\frac{\bar{\gamma}^3m^3n^{-2}}{2\int_{k}\left|v_{k}\right|^{-2}}\int_{x,y,z}\frac{(x{-}y)^2(1{-}xy{-}xz{-}yz)^2}{(1{+}x^2)(1{+}y^2)(1{+}z^2)}\frac{2{+}x^2{+}y^2{+}z^2{+}xy{+}xz{+}yz}{(1{-}xy{-}xz{-}yz)^2{+}\bar{\gamma}^2(2{+}x^2{+}y^2{+}z^2{+}xy{+}xz{+}yz)^2}.
	\end{gathered}
\end{equation}
As can be seen, the main part of the integral is collected at $|x|{\gg}|y|,|z|$ and $|y|{\gg}|x|,|z|$. Instead of taking into account both regions, we can take the first one twice. Recall also that according to $n{=}\int_{k}1$, the integral over the momenta is carried out along a circle of radius $\pi n$ (or $\pi n/m$ for dimensionless variables).
\begin{equation}
	\begin{gathered}
		\frac{1}{|\tau_{\textsf{a}}|}\approx\frac{\bar{\gamma}^3m^3n^{-2}}{\int_{k}\left|v_{k}\right|^{-2}}\int_{x,y,z}\frac{x^4(y{+}z)^2}{(1{+}x^2)(1{+}y^2)(1{+}z^2)}\frac{1}{(y{+}z)^2{+}\bar{\gamma}^2x^2}=\frac{2\bar{\gamma}^3m^4}{n^2(2\pi)^3}\int\limits_{-\pi n/m}^{\pi n/m}\frac{x^4}{(1{+}x^2)}\frac{2\pi^2\,dx}{2{+}\bar{\gamma}|x|}.
	\end{gathered}
\end{equation}
The integrals over $y$ and $z$ do not diverge at infinity and therefore, in the leading order, are independent of the cutoff $\pi n/m$. The integral over $x$, on the contrary, is collected at large momenta.
\begin{equation}
	\begin{gathered}
		\frac{1}{|\tau_{\textsf{a}}|}\approx\frac{\bar{\gamma}^3m^4}{2\pi n^2}\int\limits_{-\pi n/m}^{\pi n/m}\frac{x^2\,dx}{2{+}\bar{\gamma}|x|}\approx\begin{cases}
        \pi^2\bar{\gamma}^3nm/6, &\bar{\gamma}n/m\ll1,\\
		\pi\bar{\gamma}^2 m^2/2,&\bar{\gamma}n/m\gg1.
	\end{cases}
	\end{gathered}
\end{equation}


\twocolumngrid

\bibliography{Lindblad-ref}

\begin{thebibliography}{91}
\bibitem [{els()}]{elsewhere}%
  \BibitemOpen
  \bibinfo {note} {To be published elsewhere.}\BibitemShut {Stop}%
\end{thebibliography}

\end{document}